\documentclass[aps,prx,reprint,showpacs,superscriptaddress,twocolumn, longbibliography]{revtex4-2}

\usepackage[utf8]{inputenc}       
\usepackage[T1]{fontenc}

\usepackage{graphicx}             
\usepackage{dcolumn}              
\usepackage{multirow}             
\usepackage{xcolor}               
\usepackage{lipsum}
\usepackage{enumitem} 
 
\usepackage{amsmath, amssymb, amsthm, mathtools, bm, bbm, dsfont}
\theoremstyle{definition}

\usepackage{empheq, cases}

\usepackage[colorlinks=true,
            linkcolor=blue!50!black, 
            citecolor=red,
            urlcolor=blue,
            pdfauthor={},
            pdftitle={},
            pdfsubject={},
            pdfkeywords={}]{hyperref}

\usepackage{tikz}
\usepackage{pgfplots}
\usepackage{tikz-3dplot}
\pgfplotsset{compat=newest}
\usepgfplotslibrary{colormaps, patchplots}
\usetikzlibrary{arrows.meta, decorations.text, decorations.markings,
                shapes.symbols, shapes.arrows, intersections,
                mindmap, trees, backgrounds, fadings, shadows,
                positioning, calc, patterns, snakes}

\definecolor{c1}{RGB}{219,68,56} 
\definecolor{c2}{RGB}{74,91,163}
\definecolor{c3}{RGB}{105,182,106}
\definecolor{c4}{RGB}{237,145,33}
\definecolor{c5}{RGB}{176,48,96}

\theoremstyle{definition}

\newcolumntype{C}[1]{>{\centering\arraybackslash}p{#1}}

\begin{document}

\title{Continuous Unitary Designs for Universally Robust Quantum Control}
 
\author{Xiaodong Yang}
\affiliation{Institute of Quantum Precision Measurement, State Key Laboratory of Radio Frequency Heterogeneous Integration, College of Physics and Optoelectronic Engineering, Shenzhen University, Shenzhen 518060, China}
\affiliation{Quantum Science Center of Guangdong-Hong Kong-Macao Greater Bay Area (Guangdong), Shenzhen 518045, China}

\author{Jiaqing Leng}
\affiliation{Institute of Quantum Precision Measurement, State Key Laboratory of Radio Frequency Heterogeneous Integration, College of Physics and Optoelectronic Engineering, Shenzhen University, Shenzhen 518060, China}

 \author{Jun Li}
\email{lijunquantum@szu.edu.cn}
\affiliation{Institute of Quantum Precision Measurement, State Key Laboratory of Radio Frequency Heterogeneous Integration, College of Physics and Optoelectronic Engineering, Shenzhen University, Shenzhen 518060, China}
\affiliation{Quantum Science Center of Guangdong-Hong Kong-Macao Greater Bay Area (Guangdong), Shenzhen 518045, China}
 
\begin{abstract}
Unitary designs are unitary ensembles   that emulate      Haar-random unitary statistics. They provide a vital tool   for studying  quantum randomness and have found broad applications in quantum technologies. 
%They underpin  diverse foundational studies in complex quantum systems, as well as  practical tasks like quantum benchmarking. 
However, existing research has   focused on discrete ensembles,   despite that many   physical processes, such as in quantum chaos, thermalization, and control, naturally involve continuous ensembles generated from continuous time-evolution.  Here  we initial the study of continuous unitary designs, addressing     fundamental questions about their   construction  and   practical utility. For single-qubit system, we construct explicit unitary 1-design paths  from spherical 2-design curves and   Hopf fibration theory.   For  arbitrary dimensions, we develop two systematic construction frameworks, one based on topological bundle theory of the unitary group and the other based on the Heisenberg-Weyl group.  On the practical front,   our unitary design paths provide   analytical solutions to   universally robust quantum control. Simulations show they outperform  conventional pulse techniques  in mitigating arbitrary unknown static noises, demonstrating immediate utility for quantum engineering. Extending unitary designs to the continuous domain  not only introduces powerful geometric and topological tools that complement conventional combinatorial and group-theoretic methods, but also enhances experimental feasibility over discrete counterparts which usually involve instantaneous pulses. As an outlook, we anticipate that this work will  pave the way for using continuous unitary designs to  explore complex quantum   dynamics and devise    quantum information protocols.
\end{abstract}

\maketitle

\section{Introduction}

This paper   develops analytical methods for constructing  \emph{continuous  unitary  designs}. We define a  continuous  unitary $t$-design path  as a unitary ensemble $\mathcal{E} = \{U(s)\}_{s\in[0,1]} $ that corresponds to a path  in the unitary group and satisfies  
\begin{equation}
 \int_{[0,1]} f(U(s)) \, ds
   = \int f(U)  dU 	
   \label{unitary-1design-path}
\end{equation}
  for any   polynomial  $f$  up to degree   $t$, where  $dU$ denotes the normalized Haar measure.  These continuous designs generalize their discrete counterparts and enable important applications such as robust quantum control.

Randomness is fundamental to quantum processes and comprises a powerful resource for quantum technologies including algorithms, cryptography, and control. 
Harnessing this resource   relies critically  on the ability to implement random unitary operators. 
However, generating truly   random unitaries by uniform sampling from the unitary group under Haar measure is inherently difficult,  as      most unitaries require exponential time to implement \cite{NielsenChuang2000}. To address this difficulty,   researchers introduced  \emph{unitary designs}, which are efficiently implementable unitary ensembles capable of emulating the statistical properties of true randomness \cite{emerson2003pseudo,Eisert07,Dankert2009}.    Formally,   an ensemble   forms a unitary $t$-design if its  first  $t$ statistical moments match   the uniform Haar distribution \cite{PhysRevLett.116.200501}. Substantial research has explored generating such designs (exactly or approximately) via random quantum circuits \cite{Harrow09,PhysRevLett.104.250501,Horodecki16,PhysRevLett.116.170502,PhysRevLett.134.180404,harrow2023approximate,PRXQuantum.5.040349,PRXQuantum.5.040344,PhysRevX.15.021022} or other methods \cite{PhysRevX.7.021006,PhysRevLett.123.030502,PhysRevA.97.022333,PhysRevX.7.041015,haferkamp2023efficient}.

The capability to emulate Haar-random properties makes unitary designs  exceptionally valuable   both in theory and applications. Theoretically, they provide a  powerful tool for studying random behaviors in many-body systems,  advancing our understanding of fundamental phenomena including quantum    chaos \cite{Yoshida17,choi2023preparing,PhysRevLett.134.180403}, thermalization \cite{PhysRevX.15.011031,liu2018entanglement}, entanglement evolution \cite{PhysRevX.7.031016,PhysRevLett.120.130502},   and complexity growth \cite{haferkamp2022linear,PRXQuantum.2.030316}.
% complex phenomena in quantum thermodynamics  \cite{popescu2006entanglement,PhysRevE.79.061103,PhysRevB.108.104317} and many-body physics \cite{choi2023preparing,PhysRevLett.131.110601}.
Practically, unitary designs have become an indispensable tool in      quantum information processing. For example, the single-qubit   Pauli group   forms a     unitary 1-design that possesses   universal decoupling property \cite{PhysRevLett.82.2417,PhysRevLett.102.080501}. The multiqubit Clifford group  is a      unitary 2-design that plays a vital role  for quantum estimation, characterization, and benchmarking protocols \cite{Laflamme07,PhysRevA.80.012304,PhysRevLett.106.180504,PhysRevLett.121.170502,PRXQuantum.6.030202,PRXQuantum.2.030339}. Additionally, unitary designs have   demonstrated utility  in cryptography \cite{Winter04,lancien2020weak}, sensing \cite{PhysRevX.6.041044}, error correction \cite{bnld-2chd},  etc.  
Given their remarkable versatility, unitary designs continue to drive substantial   efforts toward their constructions  and applications \cite{mcclean2018barren,PhysRevA.101.042126,Modi21,PhysRevX.14.041059,schuster2025random,21vm-bz3t,PRXQuantum.6.010345,dy4m-gq5c,PhysRevLett.125.250501,PhysRevLett.116.040501,ketterer2020entanglement,PhysRevResearch.6.023020}.

 Despite extensive research on unitary $t$-designs,  the focus to date has predominantly been on discrete ensembles. However, many contexts naturally necessitate the consideration of continuous ensembles. A prominent example is  universally robust control (URC),   which    aims to implement quantum gates resilient to  arbitrary unknown noise.  Recently, Ref.~\cite{Poggi2024} established that  URC corresponds precisely to the condition that the controlled evolution forms a unitary 1-design. The authors thence discretized the evolution and   numerically optimized control pulses to make the   discrete-time propagators  satisfy this condition. However, numerical approaches face computational bottlenecks as system dimensionality grows exponentially   with the number of qubits.   Besides, the resulting solutions often  lack physical interpretability. This raises the question: can we  construct  analytical   continuous unitary 1-designs for   URC? Indeed, experimentally implementable robust control pulses with simple analytical forms would be particularly valuable  for practical applications.
Beyond quantum control, in studying dynamical and statistical phenomena such as  quantum chaos, thermalization, and ergodicity that arise in quantum many-body   systems,  it is also natural to consider continuous ensembles generated  from continuous-time evolution \cite{PhysRevX.15.011031,PhysRevA.101.042126,PhysRevX.14.041059}.  
%Extending design theory to continuous regime creates opportunities to  employ powerful mathematical tools from analysis, geometry, and topology, complementing conventional combinatorial and group-theoretic approaches. Our work  presents a step  toward realizing the potential of this expanded   perspective. 

A further motivation stems from recent developments in   classical   continuous spherical designs.  As unitary designs trace their origin to spherical designs \cite{adam2013applications}, a comparative study of their progress can be particularly fruitful. A spherical $t$-design  is a finite subset of points on the $d$-sphere $\mathbb{S}^d$ such that  any polynomial of degree at most $t$ has the same average on this set as on the entire sphere. Since their introduction in  the  late 1970s \cite{Delsarte1977}, spherical designs have been intensively studied, yet almost exclusively as discrete sets. Recently, mathematicians have paid attention to   continuous versions \cite{Ehler2023,Ehler25,Lindblad1,Lindblad2}.  Ref.~\cite{Ehler2023} defines a   \emph{spherical $t$-design curve} as   a closed   curve $\gamma:[0,1]\to \mathbb{S}^d$ of arc length $\ell(\gamma)$ satisfying
$\frac{1}{\ell(\gamma)}\int_\gamma f  = \int_{\mathbb{S}^d} f 
$ for all  polynomials $f$ in $d+1$ variables of degree at most $t$. The fundamental      question is hence   how to construct curves with such design properties. This leads to novel   construction methods for spherical designs, such as approaches based on   Hopf fibration    \cite{Lindblad1,Lindblad2}. Given the close relationship between unitary and spherical designs, it is   natural and interesting to extend these concepts    and methodologies to the study of continuous unitary designs.

Motivated by both the demand in quantum applications and the advances in classical spherical design curves, we initiate the study of continuous unitary designs. In this work, we focus on two fundamental aspects: construction and application. First, how to analytically construct unitary 1-design paths, as defined in Eq. (\ref{unitary-1design-path})? A natural approach is to find a continuous path that connects a discrete unitary 1-design set, while ensuring the path itself also forms a design \cite{dd2}. However,   this turns out to be  more challenging than it initially appears, especially in high-dimensional spaces. Our core idea  leverages the profound connection between unitary designs and spherical designs.   For a two-level system, the well-known equivalence between the manifold $\mathbb{SU}(2)$ and the 3-sphere $\mathbb{S}^3$ permits the direct conversion of a spherical 2-design curve   into a unitary 1-design path. However, this   correspondence  does not extend directly to higher dimensions since $\mathbb{SU}(d)$ is not isomorphic to a sphere for $d > 2$. Despite this, we can   exploit the fiber bundle structure of $\mathbb{SU}(d)$, which is based over odd-dimensional spheres \cite{husemoller1966fibre,Frankel2011}. This enables us to   employ a lifting technique developed from \cite{Lindblad1, Lindblad2} to construct continuous unitary designs for arbitrary dimension $d$.

Second, regarding practical utility, continuous unitary design is of significance for tackling the noise challenge in near-term quantum technologies--specifically, the problem of URC as introduced earlier.
 In current   quantum devices, noise is unavoidable and often arises from multiple concurrent sources \cite{Wallman20,reagor2018demonstration,schafer2018fast}. As   systems scale up,  fully characterizing all the noise components in the exponentially large Liouvillian space becomes increasingly difficult \cite{PRXQuantum.6.030202}. Robust control aims to achieve high-fidelity   gates while suppressing the noise effects  without requiring full noise characterization \cite{Schirmer25}. However, widely used robust control techniques   such as dynamical decoupling \cite{dd1,dd2,PhysRevApplied.18.054075}, composite pulses \cite{Levitt86,composite1,composite2,CCP13}, and geometric evolution   optimization \cite{PhysRevLett.111.050404,PhysRevLett.125.250403,PhysRevA.99.052321,PRXQuantum.2.010341,yang2024quantum},  are usually limited to mitigating noise along specific  directions. There is thus a   need to   develop URC methods  applicable   to arbitrary noise orientations and arbitrary target gates \cite{Poggi2024,PhysRevApplied.22.014060,ding2025universally}. Our continuous unitary design framework can generate analytic solutions for URC.  Numerical simulations confirm that the resulting analytic pulses enable robust single-qubit identity and dynamical decoupling operations, offering immunity to arbitrary static noise while outperforming conventional composite pulses  and dynamical decoupling sequences. These pulses also exhibit robustness against multi-axis, slowly varying, and time-correlated noise. This work establishes a foundation for the analytical design of URCs, opening a viable route toward reliable quantum control in realistic, noisy experimental settings.

Our main results are summarized as follows:
\begin{enumerate}
\item[(1)] For a single-qubit system,   we construct     a closed continuous   unitary 1-design path of length  $\sqrt{5}\pi$, defined by
\begin{equation}
	U(\theta) = R_{\bm{n}_1}(\theta) R_{\bm{n}_2}(2\theta),  
	\label{1-design-path-1}
\end{equation}
where   $\theta = 2\pi s$ with $s\in [0, 1]$, $\bm{n}_1$ and $\bm{n}_2$ are any pair of mutually perpendicular rotational axes. This path is generalizable, allowing its endpoint to be reconfigured to match an arbitrary target operator. As an example,    a valid   unitary 1-design path yielding a  $\pi$ rotation  about the $z$-axis is  
\begin{equation}
	U(\theta(s)) = R_z(\theta(s)) R_y(2\theta(s)),   \label{1-design-path-2}
\end{equation}
where $\theta(s):[0,1]\to [0,3\pi]$ is a continuous piecewise-linear function:
\begin{numcases}{\theta(s)=}
  4\pi s, &   $s \in [0,1/4]$, \nonumber \\
  2\pi s + \pi/2, &   $s \in [1/4,3/4]$, \nonumber\\
  4\pi s - \pi, &   $s \in [3/4,1]$. \nonumber
\end{numcases}
A notable feature of the path $U(\theta)$ in Eq. (\ref{1-design-path-1}) is that any finite equiangular sampling of  $N\ge 4$    points     $\{ \theta_k = 2\pi k/N \}_{k=0}^{N-1}$ of the path  still forms a unitary 1-design. This property is advantageous for experimental pulse controls with finite time resolution. In contrast, Eq. (\ref{1-design-path-2}) does not possess this property.
\item[(2)] For higher-dimensional systems, we present two general construction methods. The first leverages the topological structure of the $\mathbb{SU}(d)$ group. Using the fiber bundle $\mathbb{SU}(d-1) \hookrightarrow \mathbb{SU}(d) \to \mathbb{S}^{2d-1}$, we develop an inductive procedure to construct a unitary 1-design path in $\mathbb{SU}(d)$ from a corresponding path in $\mathbb{SU}(d-1)$ and a spherical 2-design curve on $\mathbb{S}^{2d-1}$. The second method is based on the a discrete unitary 1-design  Heisenberg-Weyl group. We construct a path that connects its elements and simultaneously forms a unitary 1-design.
\item[(3)] We   derive  analytic  URC pulse from the constructed unitary 1-design paths and verify their performance through numerical simulations. As an illustrative example, we focus  primarily on single-qubit control based on the path specified in Eq. (\ref{1-design-path-1}). It is important to emphasize that this equation does not represent applying two separate quantum gates, but rather defines a controlled evolution path. Our objective is   to inversely determine a corresponding control pulse that generates this   path exactly. Using inverse engineering, we obtain the explicit pulse form as follows:
\begin{equation*}
\hspace{1cm}
\bm{u}(t)  = \frac{\Omega}{\sqrt{5}} \left(-2\sin\left(\frac{\Omega  }{\sqrt{5}} t \right), 2\cos\left(\frac{\Omega }{\sqrt{5}}  t\right), 1 \right),
\end{equation*}
where  $t \in [0,  2\pi \sqrt{5}/\Omega]$, and $\Omega$ is the maximum Rabi power that is allowed under experimental constraints. We compared its performance with conventional composite pulses like CORPSE and BB1, and found that URC can resist static or slowly varying noise from any direction.  Furthermore, repeated applications of this URC pulse   form a dynamical decoupling sequence applicable for quantum memory. Our simulations demonstrate that this sequence outperforms conventional sequences like CPMG and XY4 in preserving quantum memory under static noise of unknown amplitude acting along arbitrary spatial directions. 
%\begin{equation}\label{urc_gate}
%\begin{split}
%\bm{u}(t) &= \frac{\Omega}{\sqrt{5}} \Big[ \bm{n}_1 + 2 \cos(\Omega t) \bm{n}_2 \\
%&\quad + 2 \sin(\Omega t) (\bm{n}_1 \times \bm{n}_2) \Big],
%\end{split}
%\end{equation}
\end{enumerate}

\section{Preliminaries}
\label{section-II}
This section covers the basic notations and notions,  and some known  results that serve as the basis for this work. 

\subsection{Unitary Design}

%{\color{red}remove terminology of $t$-fold twirling. definition modify to sum equal to integral}

Let $\mathcal{H}\cong \mathbb{C}^d $ denote the $d$-dimensional Hilbert space and $\mathbb{U}(d)$   the group of unitary operators.
A unitary \(t\)-design can be   characterized  through   $t$-fold quantum twirling  channels \cite{PRXQuantum.6.030202}. For a finite ensemble   \(\mathcal{E} = \{U_k\}_{k=1}^K \subset \mathbb{U}(d) \) with uniform weights,   the   \(t\)-fold twirling channel \(\Phi_{\mathcal{E}}^{(t)}  \) acts on   any linear  operator $A$ of the  $t$-fold Hilbert space $  \mathcal{H}^{\otimes t} $ as
\[
\Phi_{\mathcal{E}}^{(t)}(A) = \frac{1}{K} \sum_{k=1}^K U_k^{\otimes t} A (U_k^{\otimes t})^\dagger.
\]
This definition can be extended to continuous   ensembles by replacing the summation with integration. Let   $\Phi_\text{Haar}^{(t)}$ denote  the $t$-fold Haar-random twirling channel  for  $\mathbb{U}(d)$   with     normalized   Haar measure,
%\[
%\Phi_{\mathcal{E}}^{(t)}(O) = \int_{\mathcal{E}} dU  U^{\otimes t} O (U^{\otimes t})^\dagger.
%\]
then  \(\mathcal{E}\) forms a unitary \(t\)-design if and only if   
\[
\Phi_{\mathcal{E}}^{(t)}(A) = \Phi_\text{Haar}^{(t)}(A) 
\]
for all \(A\).
This   formulation establishes statistical indistinguishability of \(\mathcal{E}\)  from Haar unitaries up to $t$th moment. 
The definition can also be expressed in terms of Pauli operators, which form a complete operator basis. Concretely, 
$\mathcal{E}$ constitutes a unitary $t$-design iff $
\Phi_{\mathcal{E}}^{(t)}(P) = \Phi_{\text{Haar}}^{(t)}(P)$
for all $P \in \{I, X, Y, Z\}^{\otimes t}$, 
where $X$, $Y$, and $Z$ denote the Pauli operators 
$\sigma_x$, $\sigma_y$, and $\sigma_z$, respectively, and $I$ represents the identity matrix. For $t=1$, the Haar ensemble channel corresponds to a depolarizing map:
\[
\Phi_{\mathcal{E}}^{(1)}(\rho) = \int_{\mathbb{U}(d)} U \rho U^\dagger  dU = \frac{\operatorname{Tr}(\rho)}{d}I,
\]
for any Hermitian operator $\rho$. Therefore, an ensemble forms a unitary 1-design iff  it averages every traceless Pauli operator to zero.

%with equivalence to polynomial moment matching following since $U^{\otimes t}$ generates all degree-$t$ polynomials in $U$ and $U^\dagger$.  

The unitary design property is invariant under     global phase transformations. That is, if $\{U_k\}$ forms a unitary $t$-design, then so does $\{e^{i \theta_k} U_k\}$ for any   choice of phases $\theta_k$. Since the elements of the   special unitary group $\mathbb{SU}(d) $ differ from those of $\mathbb{U}(d)$ only by   global phases,   we     restrict   to    $\mathbb{SU}(d)$ in this work without loss of generality. Furthermore, a unitary $t$-design is automatically a $t'$-design for $t'< t$, and remains invariant   under left multiplication,  right multiplication, or conjugation  by any fixed   $V\in \mathbb{SU}(d)$, reflecting Haar measure invariance.

The $N$-qubit Pauli group defined by $\mathcal{P}_N := \{\pm 1, \pm i\}\cdot \{I,X,Y,Z\}^{\otimes N}$ is   known for being a  unitary 1-design.    For example,    the set $\{I, iX, iY, iZ\}  $ is a   unitary 1-design in $\mathbb{SU}(2)$. The $N$-qubit Clifford group, defined as the normalizer of the Pauli group, constitutes a unitary 2-design and is known to be a 3-design but not a 4-design \cite{PhysRevA.96.062336}.

A powerful diagnostic tool for verifying unitary design properties is the $t$th frame potential \cite{Yoshida17}, defined as:
\begin{equation}
    \mathcal{F}^{(t)}_{\mathcal{E}}  = \frac{1}{K^2} \sum_{k,j=1}^K   \left|\operatorname{Tr}(U_k^\dagger U_j)\right|^{2t}.
\end{equation}
This quantity provides a lower bound for the Haar frame potential $\mathcal{F}^{(t)}(\mathcal{E}) \ge \mathcal{F}^{(t)}_\text{Haar}=t!$, with equality iff the ensemble forms a $t$-design. In particular,   $\mathcal{F}^{(1)}  = 1$ characterizes unitary 1-designs. The frame potential serves as an efficiently computable measure for characterizing the randomness of unitary ensembles and quantifying their closeness to the Haar measure.

While most research focuses on discrete unitary designs, this work investigates continuous unitary design paths. We say an ensemble 
\(\mathcal{E} = \{U(s)\}_{s \in [0,1]}\) is a \emph{path} if it continuously maps 
\(s \in [0,1]\) to \(U(s) \in \mathbb{SU}(d)\). The path connects two unitaries \(U(0)\) and \(U(1)\)  as its endpoints. When \(U(0) = U(1)\), 
the path is said to be closed (or a loop). The ensemble  $\mathcal{E}$ forms a continuous unitary $t$-design in $\mathbb{SU}(d)$ if, for every polynomial $f$ of degree at most $t$ in the matrix elements of $U$ and $U^\dagger$,
\[
   \int_{[0,1]} f(U(s)) \, d\mu(s)
   = \int_{\mathbb{SU}(d)} f(U)  dU, 
\]
where $\mu$ is uniform measure on $[0,1]$ and $dU$ is normalized Haar measure.  A natural approach to constructing  such   paths is to appropriately connect points from an already known discrete unitary design. However, this seems to be quite a   non-trivial task. This work is devoted to developing methods to realize this construction principle.

\subsection{Spherical Design}
Define $\mathbb{S}^d = \{x\in \mathbb{R}^{d+1}:\|x\| =\sum_{k=1}^{d+1} x_k^2 =1\}$ to be the $d$-sphere of the $(d+1)$-dimensional Euclidean space $\mathbb{R}^{d+1}$.
A spherical design is a configuration of points that are evenly distributed on the sphere.  In mathematical definition \cite{Delsarte1977}, a finite set of points $X$ on $\mathbb{S}^d$ is a spherical $t$-design if for every polynomial $f$ in $d+1$ variables of degree $t$, there is
\begin{equation}
	\frac{1}{| X |}\sum_{x\in X}f(x) =   \int_{\mathbb{S}^d} f(x) d\mu(x),
\end{equation}
where $\mu$ denotes the standard uniform measure on $\mathbb{S}^d$, normalized so that $\int_{\mathbb{S}^d} 1 d\mu = 1$. 

Some basic facts about spherical designs are given as follows. A spherical $t$-design  is automatically a  $(t-1)$-design. The property of being a $t$-design is invariant under orthogonal transformations due to the rotational symmetry of $\mu$. Furthermore, the disjoint union of spherical $t$-designs is itself a $t$-design. For a set $X \subset \mathbb{S}^d $, it is a   1-design  iff  $ \frac{1}{| X |}\sum_{x\in X} x = 0$,
and it is a 2-design iff $  \frac{1}{| X |}\sum_{x\in X} x^T  x  = I_d/d$, where  $I_d$ denotes the $d$-dimensional identity matrix.

Constructing spherical $t$-designs is generally challenging.  There exists a rich field of research on analytic and numerical methods for  their construction \cite{Bannai09,BRV13}. The most interesting $t$-designs are those of minimal
cardinality, called tight designs, but are very rare. Tight $1$-designs and $2$-designs are completely classified. A tight  1-design consists of any pair of antipodal points $x$ and $-x$ on $\mathbb{S}^d$. A tight $2$-design comprises the $d+2$ vertices of a regular simplex inscribed in $\mathbb{S}^d$.  Another basic fact is for $\mathbb{S}^1$, a  tight $t$-design   is provided by a regular $(t+1)$-gon:
\begin{center}
\includegraphics[width=0.85\linewidth]{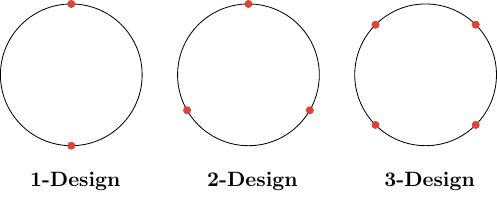}
\end{center}
These tight designs are unique up to  orthogonal transformations.

Spherical designs are normally studied in the discrete setting. 
Recently, Ref. \cite{Ehler2023}  introduced a continuous analogue, defining a spherical $t$-design curve to be  a continuous, piecewise smooth,
closed curve $\gamma:[0,1]\to \mathbb{S}^d $ with $\gamma(0)=\gamma(1)$, such that for every polynomial $f$   of degree at most $t$:
\begin{equation}
\frac{1}{\ell(\gamma)}\int_\gamma f(\gamma(s)) |\dot \gamma(s)| ds  = \int_{\mathbb{S}^d} f(x) d\mu(x),
\end{equation}
where $\dot \gamma:= d \gamma /d s $ and $\ell(\gamma) := \int_0^1 |\dot \gamma(s)|ds$ denotes the arc-length of $\gamma$.    Several initial approaches to constructing such curves have been explored, such as using graph theory, topological methods, and numerical optimization \cite{Ehler2023, Ehler25,Lindblad1}. For example, the following curve
\begin{equation}
	\gamma(s) = \frac{1}{\sqrt{d}} (\cos(2\pi s),\sin(2\pi s),..., \cos( 2\pi ds), \sin( 2\pi ds)) \nonumber
\end{equation}
forms a spherical $2$-design curve on $\mathbb{S}^{2d-1}$. This example will be   relevant to our construction of continuous unitary 1-designs later in this work.
Despite a handful of explicit examples now being known, as noted in Ref. \cite{Ehler25},   the general construction of such curves remains challenging, both numerically and analytically.

\section{Single-qubit}
\label{section-III}
\subsection{Closed Unitary 1-design Path}
We begin with the simplest single-qubit system of dimension   $d=2$. Every unitary operator in $\mathbb{SU}(2)$ corresponds to a rotational gate on the Bloch sphere, expressed as $R_{\bm{n}}(\theta) = \exp(-i \theta \sigma_{\bm{n}}/2)$, where $\theta \in [0,2\pi]$ is the rotational angle,   $\bm{n} = (n_x,n_y,n_z)$ is a   unit vector in $\mathbb{R}^3$ specifying the rotational axis, and   $\bm{\sigma} = (\sigma_x, \sigma_y,\sigma_z)$ denotes the vector of   Pauli operators. As a first result  of this work, we construct a unitary  1-design path in $\mathbb{SU} (2)$  given by
\begin{equation}
	U(\theta) = R_{\bm{n}_1}(\theta) R_{\bm{n}_2}(2\theta),  \label{1-design}
\end{equation}
where $\bm{n}_1 \perp \bm{n}_2$ and $\theta =2\pi s $ with $s\in [0, 1]$.  This forms a closed path     based at identity, since $U(0) = U(2\pi) = I$. 

We now explain how this path     constitutes  a unitary  1-design. Let us first consider the specific case with $\bm{n}_1 = (0,0,1)$ and  $\bm{n}_2 = (-1,-1,0)/\sqrt{2}$, so that the unitary path becomes  $U(\theta) =   e^{-i \theta \sigma_z /2} e^{i 2 \theta \sigma_{\pi/4}/2}$, where $\sigma_{\pi/4} \equiv (\sigma_x+\sigma_y)/\sqrt{2}$ denotes the operator along the direction $\pi/4$ in the $xy$-plane. A crucial observation is that this path intersects the single-qubit Pauli group (up to a phase factor) at the points:  $U(0) =  I$, $U(\pi/2) =  i Y$, $U(\pi) =  i Z$, and $U(3\pi/2) = iX$. 
Moreover, for  any $\varphi \in [0,\pi/2)$, consider the discrete set $\mathbf{U}_\varphi = \{U(\varphi + k\pi/2) \}_{k=0}^3$. Clearly, this set can be expressed as the coset  $\mathbf{U}_\varphi = e^{-i \varphi \sigma_z /2} \mathbf{U}_0 e^{ i 2\varphi \sigma_{\pi/4} /2} $. Since $\mathbf{U}_0 = \{ I, i\sigma_y, i\sigma_z, i\sigma_x\}$ is a unitary 1-design and  the 1-design property is invariant under either left or right multiplication by a fixed unitary, it follows that   $\mathbf{U}_\varphi$ is also a unitary 1-design for every $\varphi$. The union of these 1-designs  $\bigcup_{\varphi \in [0,\pi/2)} \mathbf{U}_\varphi$  thus gives our   continuous unitary 1-design path. 

More generally, the construction remains valid for any pair of mutually perpendicular axes.  To see this,  consider arbitrary $\bm{n}_1$ and $\bm{n}_2$ satisfying $\bm{n}_1 \perp \bm{n}_2$. One may first apply a rotation that maps the $z$-axis to the direction of $\bm{n}_1$, thereby transforming $\sigma_z$ into $\sigma_{\bm{n}_1}$. Under the same rotation,  the original axis corresponding to $\sigma_{\pi/4}$
  is mapped to a new axis lying in the plane perpendicular to $\bm{n}_1$. A subsequent   rotation  about $\bm{n}_1$ can then align this new  axis with $\bm{n}_2$ and meanwhile keeps   $ \bm{n}_1$  unchanged. The resulting path thus retains the form of Eq. (\ref{1-design}) and  constitutes  a continuous unitary 1-design.

\begin{figure}[t]
\begin{center}
\includegraphics[width=\linewidth]{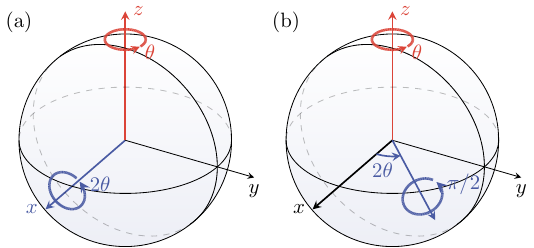}
\caption{Two examples of unitary 1-design paths constructed from composition  of   rotational gates that are controlled by the same angular variable $\theta$. (a) $U(\theta) = R_z(\theta)R_x(2\theta)$: both gates employ fixed rotational axes and varying rotational angles. (b) $U(\theta) = R_z(\theta)R_{2\theta}(\pi/2)$: one gate uses a fixed rotational axis and varying angle, while the other employs a varying rotational axis and fixed angle.}	
\label{example}
\end{center}	
\end{figure}

We provide another   unitary 1-design path of similar form
 \begin{equation}
 	U(\theta)  = R_{\bm{n}}(\theta) R_{\bm{n}_\perp(2\theta)}(\pi/2), \label{1-design-2}
 \end{equation} 
 where $\theta \in [0, 2\pi]$,  $\bm{n}$ is an arbitrary rotational axis, and $\bm{n}_\perp(2\theta)$ represents  an axis that starts from a vector  $\bm{n}_\perp$   perpendicular to $\bm{n}$ and rotates about $\bm{n}$ with angular parameter  $2\theta$. This path is   closed and  based at the operator $R_{\bm{n}_\perp }(\pi/2)$. Unlike the previous construction, here one of the rotational gates has a fixed rotational angle $\pi/2$ but a varying  rotational axis, as illustrated in Fig. \ref{example}.  
 
% Let $\bm{n}$ be an arbitrary vector, let $\bm{n}_\perp$ be an arbitrarily chosen vector that lies in the plane perpendicular to $\bm{n}$, denote $\bm{n}_\perp(\alpha)$ to be vector obtained by rotating $\bm{n}_\perp$ about $\bm{n}$-axis by angle $\alpha$,
% \begin{equation}
% 	U(\theta)  = R_{\bm{n}}(\theta) R_{\bm{n}_\perp(2\theta)}(\pi/2),
% 	\label{1-design-2}
% \end{equation}
% where $\bm{n}_\perp(\alpha)$ represents an axis in the plane perpendicular to $\bm{n}$ whose direction varies with $\alpha$.

\subsection{Continuous Unitary 1-design on $\mathbb{SU} (2)$ and Spherical 2-design curve on $\mathbb{S}^3$} 
To explain how unitary 1-design paths are constructed, we first highlight the close relationship between continuous unitary 1-designs and spherical 2-design curves. 
We begin by   identifying the group manifold of $\mathbb{SU} (2)$ with the 3-sphere $\mathbb{S}^3$. Any unitary operator  in $  \mathbb{SU} (2)$ can be expressed in the form  
\begin{equation}
U=\begin{pmatrix}  
z_1 &   z_2 \\  
-\bar z_2  & \bar z_1   
\label{mat:unitary}
\end{pmatrix}, \quad (z_1, z_2) \in \mathbb{C}^2, \nonumber	
\end{equation}
where $z = (z_1, z_2) $ satisfies the normalization condition $|z_1|^2 + |z_2|^2 = 1$. Furthermore, $\mathbb{C}^2$ is isomorphic to $\mathbb{R}^4$ via the mapping $z=(z_1,z_2) = (x_1 + i x_2,   x_3 + i x_4) \leftrightarrow x=(x_1, x_2, x_3, x_4) $. The normalization condition then  becomes $\left\|x \right\|=1$, which defines the unit     3-sphere  $\mathbb{S}^3$ embedded in $\mathbb{R}^4$.  This establishes a one-to-one correspondence  between  $\mathbb{SU} (2)$ and   $\mathbb{S}^3$, under which the Haar measure on $\mathbb{SU} (2) $ is mapped to the uniform surface measure on $\mathbb{S}^3$.  

Under the above correspondence,  one has that for any  traceless Hermitian operator $V$, the matrix   elements  of   $U V U^\dag  $ are quadratic polynomials  in the coordinates $x$ of $U$, with coefficients depending on  $V$. A spherical $2$-design on $\mathbb{S}^3$ can integrate these quadratic polynomials  exactly to match their   average   over the entire sphere. In this manner, it   induces a unitary 1-design in  $\mathbb{SU} (2)$. These arguments   are presented in more detail in Appendix \ref{appendix_A}. The converse, however, does not hold. A unitary 1-design does not necessarily yield  a spherical 2-design, as we will see later.

%Spherical 2-design curves can be constructed by forming suitable continuous connections of spherical 2-design points.   Consider  the minimal  spherical 2-design    on $\mathbb{S}^3$, given by   the five vertices of a regular 4-simplex \cite{Delsarte1977}. A concrete example of a spherical 2-design curve is given in Ref. \cite{Ehler25} as $\varsigma(\theta) = (\cos  \theta,    \sin  \theta, \cos 2\theta ,    \sin 2\theta)/\sqrt{2}$,  which passes through the  vertices at   $\{\varsigma(2\pi k/5)\}_{k=0}^4$. This curve  is not unique. Another  similar  curve  also connects these vertices is $
%	\xi(\theta) = (\cos  \theta,    \sin  \theta, \cos 3\theta ,    -\sin 3\theta)/\sqrt{2} $, whose relevance will be explained later.  The design property of   $\varsigma$ and $\xi $ can be verified by checking the following sufficient conditions for a curve $\gamma$ to be a 2-design: (i)  $\frac{1}{\ell(\gamma)} \int_{\gamma} x_k = 0$ and (ii)  $  \frac{1}{\ell(\gamma)} \int_{\gamma} x_k x_j = \delta_{kj}/4$ for $k,j=1,2,3,4$ \cite{supplement}.

\begin{figure} 
\begin{center}
\includegraphics[width=0.7\linewidth]{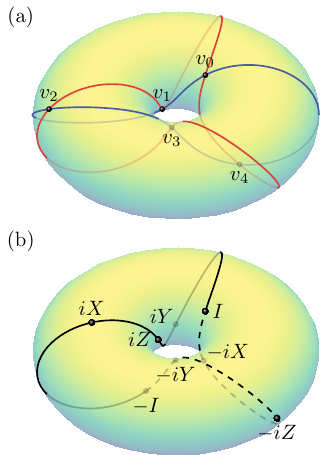}
\caption{Visualization of the continuous designs via stereographic  projection  from $\mathbb{SU} (2) \cong \mathbb{S}^3 \subset \mathbb{R}^4$ to $\mathbb{R}^3$. (a) Two continuous spherical 2-design curves  $\gamma_\phi$ and $\xi_\phi$ with    $\phi=-\pi/4$. They both pass  through   the  same set of points   $\{v_0,...,v_4\}$ which correspond to the minimal discrete spherical 2-design  on $\mathbb{S}^3$. They are also unitary 1-design paths. (b) A shorter unitary 1-design path $ \tilde \gamma_\phi$ is  constructed by taking half of   $ \gamma_\phi$ and identifying its endpoints $I$ and $-I$. }
\label{design-curve}
\end{center}	
\end{figure}

Therefore, constructing spherical 2-design curves provides a straightforward means to obtain continuous unitary 1-designs.
According to Ref.~\cite{Ehler2023}, such curves can be generated by continuously connecting the points of a discrete spherical 2-design.
The minimal  spherical 2-design    on $\mathbb{S}^3$ is given by  the five vertices of a regular 4-simplex.   Ref. \cite{Ehler25}   introduced the   curve 
\begin{equation}
\xi(\theta) = \frac{1}{\sqrt{2}}(\cos  \theta,    \sin  \theta, \cos 2\theta ,  \sin 2\theta),  	
\label{xi}
\end{equation} 
where $\theta = 2\pi s$ with $s \in  [0,1]$, and showed that it continuously interpolates the minimal 2-design set and is itself a spherical 2-design. The points     $\{\xi(2\pi k/5)\}_{k=0}^4$ on   $\xi$ correspond  to the vertices   of the simplex, as verified by checking that their pairwise inner products are all $-1/4$.   This construction, however, is not unique. Here, we are   interested in  another curve with similar behavior,   
\begin{equation}
\gamma(\theta) = \frac{1}{\sqrt{2}}(\cos  \theta,    \sin  \theta, \cos 3\theta ,   - \sin 3\theta) ,
\label{gamma}
\end{equation} 
which  also connects the same set of vertices as $\xi$. By direct calculation, the two curves intersect precisely at these vertices
\[  \gamma \cap \xi = \{\gamma(2\pi k/5) \}_{k=0}^4  = \{\xi(2\pi k/5) \}_{k=0}^4 = \bm{v}(\Delta^4) ,  \]
where we use $\bm{v}(\Delta^4)$ to denote  the vertex set of the regular 4-simplex $\Delta^4$; see Fig. \ref{design-curve}(a).
The spherical 2-design property of $\gamma$ and $\xi$ can be confirmed by verifying that both curves satisfy the following sufficient conditions:   (i)  $\frac{1}{\ell(\gamma)} \int_{\gamma} x_k = 0$ and (ii)  $  \frac{1}{\ell(\gamma)} \int_{\gamma} x_k x_j = \delta_{kj}/4$.

By using the arc length   formula, we obtain   $\ell(\xi) = \int_0^1 |\dot \xi(s)| ds = \sqrt{10}\pi$ and $\ell(\gamma) = \int_0^1 |\dot \gamma(s)| ds = 2\sqrt{5}\pi$. The curve $\xi$ is shorter,   suggesting it may be a preferable candidate for a    unitary 1-design.  However, we observe that $\gamma$ exhibits redundancy:    the point set   $\{\gamma(2\pi k/8) \}_{k=0}^7$       maps    two-to-one   onto the Pauli group 1-design (ignoring global phase), as shown in Fig. \ref{design-curve}(b). Examining the expression   for $\gamma$ in  Eq. (\ref{gamma}), we find that for any $\theta \in [0,\pi]$ there is $\gamma(\theta) = -\gamma(\theta+\pi)$, which implies that the corresponding unitary gates satisfy $U(\theta) = -U(\theta+\pi)$. Although   $U$ and $-U$ are distinct elements of $\mathbb{SU}(2)$,   they represent the same quantum gate due to the global phase invariance of quantum states. This is a consequence of the fact that $\mathbb{SU}(2)$ is a double cover of the 3-dimensional rotational group $\mathbb{SO}(3)$. To remove this redundancy, we   work in the quotient space $\mathbb{SO}(3) \cong \mathbb{SU}(2)/\{I,-I\} \cong \mathbb{S}^3/\{1,-1\}$, identifying antipodal points. The curve $\gamma$ then descends to a closed curve 
\begin{equation}
\widetilde \gamma(\theta) = \frac{1}{\sqrt{2}}(\cos  \frac{\theta}{2},    \sin  \frac{\theta}{2}, \cos \frac{3\theta}{2} ,   - \sin \frac{3\theta}{2}) ,
\label{tilde-gamma} 
\end{equation}
with  length   $\ell(\widetilde \gamma) = \ell( \gamma)/2 = \sqrt{5}\pi$, which is shorter than $\xi$. Although $\widetilde \gamma$ is not a spherical 2-design curve in $\mathbb{S}^3$, its image   in $\mathbb{SU}(2)$, given by
\[    U_{\widetilde \gamma}(\theta) = R_{-z}(\theta) R_{-\pi/2+ 2\theta}(\pi/2), \]
forms a continuous unitary 1-design, belonging to the class described by   Eq. (\ref{1-design-2}). 

\begin{table*}  
{\setlength\tabcolsep{5pt}\renewcommand{\arraystretch}{2}
\centering\footnotesize 
\begin{tabular}{|c|c|c|c|}
\hline
\multirow{3}{*}{ Spherical Curves}  &  $\displaystyle  \xi=\frac{1}{\sqrt{2}}(   e^{i\theta},     e^{i2\theta } )$    & $\displaystyle  \gamma=\frac{1}{\sqrt{2}}(   e^{i\theta},     e^{-i3\theta } )$ &  $\displaystyle  \tilde \gamma=\frac{1}{\sqrt{2}}(   e^{i\theta/2},     e^{-i3\theta/2 } )$ \\  
 &  $\displaystyle \xi'=( \cos \frac{\theta}{2}  e^{i 3\theta/2},  -\sin \frac{\theta}{2}  e^{i3\theta/2} ) $   & $ \gamma'=( \cos 2\theta  e^{-i\theta},  \sin 2\theta  e^{-i\theta} )$ &  $\tilde \gamma'=( \cos  \theta  e^{-i\theta/2},  \sin \theta  e^{-i\theta/2} ) $ \\  
 &  $\displaystyle \xi_\phi=( \cos \frac{\theta}{2}  e^{i 3\theta/2},  -\sin \frac{\theta}{2}  e^{i3\theta/2} e^{-i\phi})  $   & $ \gamma_\phi=( \cos 2\theta  e^{-i\theta},  \sin 2\theta  e^{-i\theta} e^{-i\phi})$  & $\tilde \gamma_\phi=( \cos  \theta  e^{-i\theta/2},  \sin \theta  e^{-i\theta/2} e^{-i\phi}) $ \\ \hline
\multirow{2}{*}{ Design Property } &  Spherical 2-design &  Spherical 3-design &  \multirow{2}{*}{Unitary 1-design} \\  
  &  Unitary 1-design &  Unitary 1-design &  \\ \hline
 Length  & $  \sqrt{10}\pi$  & $  2\sqrt{5}\pi$ &  $ \sqrt{5}\pi$ \\ 
\hline
\end{tabular}	
}	
\caption{A summary of the spherical curves and their design properties described in Sec. \ref{section-III}. The curves are parameterized using $\theta=2\pi s$, with $s\in[0,1]$,  and a fixed parameter $\phi\in[0,\pi]$.}\label{table1}
\end{table*}

\subsection{Hopf Fibration}

Having presented the explicit design examples, as summarized in Table. \ref{table1}, we now delve deeper into the profound geometric intuition and rationale underlying their construction through the lens of the Hopf fibration, basically following the framework established by Refs.~\cite{Lindblad1,Lindblad2}. The Hopf map, introduced by Heinz Hopf
in a seminal work in 1931 \cite{hopf1931abbildungen},  is a projection of the 3-sphere onto the 2-sphere  defined by
\begin{align}
\pi: \quad \mathbb{S}^3 & \to \mathbb{S}^2, \nonumber \\
 (z_1,z_2)  & \mapsto (|z_1|^2 - |z_2|^2,   \operatorname{Re}(2 z_1 \bar z_2),  \operatorname{Im}(2 z_1 \bar z_2)).	 \nonumber
\end{align}
%\[ \pi: \mathbb{S}^3 \to \mathbb{S}^2, \quad (z_1,z_2)  \mapsto (|z_1|^2 - |z_2|^2, 2z_1 \bar z_2, 2z_1 \bar z_2),\] 
%where the latter two components collectively represent $2z_1 \bar z_2 $. 
This is a continuous surjection with the key property that, the preimage (called \emph{fiber}) of every   point on the base sphere $\mathbb{S}^2$
  is a great circle (topologically equivalent to $\mathbb{S}^1$) in $\mathbb{S}^3$. Moreover,   fibers over distinct points  are mutually disjoint in $\mathbb{S}^3$, and their   union seamlessly reconstructs the entire $\mathbb{S}^3$. This  endows $\mathbb{S}^3$ with a    fiber bundle structure, formally denoted as  $\mathbb{S}^1  \hookrightarrow  \mathbb{S}^3 \to \mathbb{S}^2$, which effectively partitions  $\mathbb{S}^3$ into a continuous, non-intersecting family of great circles, each uniquely indexed by a point on the 2-sphere:
  \[ \mathbb{S}^3 = \bigcup_{p \in \mathbb{S}^2} \pi^{-1}(p).\]

The above decomposition naturally motivates a sampling strategy for constructing spherical design curves. One can first sample points from the base space $\mathbb{S}^2$, and then sample the corresponding circular fibers attached to these points. To obtain continuous designs, the sampling must be performed in a continuous manner. Ref.~\cite{Lindblad2} presented a concrete realization of this strategy  by showing that it is possible to select a   $\lfloor t/2 \rfloor$-design curve from $\mathbb{S}^2$ and then lift it to a $t$-design curve on $\mathbb{S}^3$. This  lifting technique is particularly helpful to understand the geometric feature of the constructed continuous designs.

\begin{figure}[b]
\begin{center}
\includegraphics[width=\linewidth]{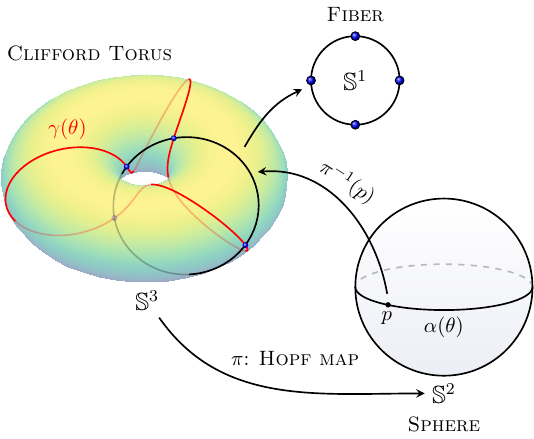}
\caption{Illustration of the construction and geometrical properties of the spherical 3-design curve $\gamma$. Hopf fibration represents the group \( \mathbb{SU}(2) \cong \mathbb{S}^3 \) as a family of circles (fibers) indexed by the base space \( \mathbb{S}^2 \). The construction begins by selecting a spherical 1-design \( \alpha  \) (a great circle)  on the base sphere. The preimage of any point $p$ on $\alpha$ under the Hopf map is a fiber (circle), and the preimage of the entire great circle $\alpha$ forms a Clifford torus in  $\mathbb{S}^3$. The curve $\gamma$ is constructed as  a   lift of $\alpha$  such that it intersects each fiber at exactly four symmetric points--forming a  spherical 3-design on that fiber. In this way,   \( \gamma \) itself constitutes a continuous spherical 3-design curve.}
\label{hopf}
\end{center}	
\end{figure}

We now explain how the spherical curve $\gamma(\theta)$ in Eq.~(\ref{gamma}) is constructed using the lifting technique. We begin by selecting a spherical 1-design curve on $\mathbb{S}^2$, typically a meridian great circle parameterized as $(0, \cos\theta, \sin\theta)$. Next, we take its fourfold cover and define the 1-design curve as $\alpha(\theta) = (0, \cos 4\theta, \sin 4\theta)$. We then lift $\alpha(\theta)$  to a curve $\gamma(\theta)  = \frac{1}{\sqrt{2}} \,(e^{i \theta}, e^{-i 3\theta})$, which satisfies $\pi(\gamma)=\alpha$.
We further clarify that $\gamma(\theta)$ constitutes a spherical 3-design curve on $\mathbb{S}^3$. For any point $p$ on the great circle, there exists $\theta_p \in [0, \pi/2]$ such that $\{\alpha(\theta_p + k\pi/2)\}_{k=0}^3$ all coincide with $p$. The fiber over $p$, explicitly written as $\pi^{-1}(p) = \left\{ \frac{1}{\sqrt{2}} \,(e^{i(\theta_p + \varphi)}, e^{-i 3\theta_p} e^{i\varphi}) : \varphi \in [0, 2\pi] \right\} \cong \mathbb{S}^1$, is intersected by $\gamma$ at $\varphi = \{k \pi/2\}$, corresponding to the vertices of a square—a spherical 3-design on the fiber. According to Ref.~\cite{Lindblad2}, $\gamma$ is thus a spherical 3-design on $\mathbb{S}^3$. In addition, we note that $\gamma$ is automatically a spherical 2-design as described in Sec. \ref{section-II}.

 The curve $\xi(\theta) =  (  e^{ i \theta },   e^{ i 2\theta })/\sqrt{2}$ in Eq. (\ref{xi})  involves a slightly more analysis. Consider the  related curve $\varsigma(\theta) =  (  e^{ i \theta },   e^{- i 2\theta })/\sqrt{2}$, which clearly   exhibits a  lifting  structure   analogous to $\xi(\theta)$: (i) its projection $\pi(\varsigma)$ covers a great circle  of $\mathbb{S}^2$, and (ii) its intersection with any fiber over that great circle corresponds to a regular triangle---a  2-design on the fiber $\mathbb{S}^1$. Hence,   $\varsigma$ forms a spherical 2-design on $\mathbb{S}^3$.  Since $\xi$ can be obtained from $\varsigma$ via the transformation $\xi = T(\varsigma)$, where $T = \operatorname{diag}(1,1,1,-1) $ represents  an inversion, it follows that $\xi$ is also a 2-design curve.

\subsection{Closed Unitary 1-design Path Based at Identity}
We are in  particular  interested  in constructing unitary 1-design paths based at the identity, as   many applications, such as quantum control, require  the evolution to start  from the identity operation.   This can be readily achieved in several ways. 

First, if $\{U(\theta):\theta\in[0,2\pi]\}$ is a unitary 1-design loop with $U(0) \ne I$, then applying $U^\dag(0)$  to the entire set yields $\{U(\theta)U^\dag(0) :\theta\in[0,2\pi]\}$, which remains   a unitary 1-design loop but now is based at identity $I$.

Second, we may     exploit  the rotational invariance of the spherical design property: any known spherical $2$-design curve gives rise to a family of equivalent curves under rotations. By applying a suitable rotation, we align the  curves to start at   $(1,0,0,0) \in \mathbb{S}^3$, corresponding to the identity gate.   Explicitly, using the rotation matrix 
\[ R =\frac{1}{\sqrt{2}} \left(\begin{matrix}
	1 & 0 & 1 & 0 \\
	0 & 1 & 0 & 1 \\
	0 & 1 & 0 & -1 \\ 
	-1 & 0 & 1 & 0
\end{matrix} \right), \quad \operatorname{det}(R) = 1,  \]
we obtain the following curves  
\begin{align}
&\xi' \equiv R(\xi) =  ( \cos \frac{\theta}{2}  e^{i3\theta/2},  -\sin \frac{ \theta}{2}  e^{i3\theta/2} ),	 \label{xip} \\
&\gamma' \equiv R(\gamma) =  ( \cos 2\theta  e^{-i\theta},  \sin 2\theta  e^{-i\theta} ), \label{gammap}  \\
&  \widetilde \gamma' \equiv R(\widetilde\gamma) =  ( \cos  \theta  e^{-i\theta/2},  \sin \theta  e^{-i\theta/2} ),   \label{tilde-gammap}  	
\end{align} 	
where $\theta = 2\pi s$ with $s \in  [0,1]$.

Third, we may alternatively employ the lifting technique from the outset to obtain similar results. For the curve corresponding to Eq. (\ref{xip}), we begin by selecting a spherical 1-design curve on $\mathbb{S}^2$ starting at the base point $(1,0,0)$. Specifically, we take the great circle $(\cos\theta, \sin\theta, 0)$ and use its threefold cover, defining the resulting 1-design curve as $ \alpha(\theta) =  (\cos 3\theta , \sin 3\theta, 0)$. Subsequently, we lift it to the following curve $\varsigma(\theta)=(\cos(3\theta/2) e^{-i\theta/2},\sin(3\theta/2) e^{-i\theta/2})$ that starts at $\varsigma(0)=(1,0,0,0)$. It is straightforward to verify that this curve satisfies $\pi(\varsigma) = \alpha$, and that its intersection with each fiber of the great circle forms a 2-design on that fiber. Consequently, $\varsigma(\theta)$ constitutes a spherical 2-design curve. Finally, we obtain the following curve $\xi_\phi$ as  
\begin{equation}
	\xi_\phi(\theta)=( \cos \frac{\theta}{2}  e^{i 3\theta/2},  -\sin \frac{\theta}{2}  e^{i3\theta/2} e^{-i\phi})
\end{equation}
from the transformation $\xi_\phi=Q(\varsigma)$ with 
\begin{equation*}
Q(\phi) = 
\begin{pmatrix}
1 & 0 & 0 & 0 \\
0 & 0 & 1 & 0 \\
0 & \cos\phi & 0 & -\sin\phi \\
0 & \sin\phi & 0 & \cos\phi
\end{pmatrix}, \quad \text{det}(Q)=1.
\end{equation*}
According to the analysis in Sec. \ref{section-II}, the set $\{\xi_\phi(\theta)\}_{\phi \in [0,\pi]}$ also constitutes a family of spherical 2-design curves.

For the curves related to Eqs. (\ref{gammap}) and (\ref{tilde-gammap}), we consider a great circle on  $\mathbb{S}^2$ obtained by rotating the equator about the $(1,0,0)$-axis by an angle $\phi \in [0,\pi]$, i.e.,  $ (\cos \theta , \sin \theta \cos \phi, \sin \theta \sin \phi) $. We see that $\phi$ plays the role of a relative phase here.  We then take its fourfold cover $ \alpha_\phi(\theta) =  (\cos 4\theta , \sin 4\theta \cos \phi, \sin  4\theta  \sin \phi)$ as a spherical 1-design curve, which is then lifted to the following curve
\begin{equation}
\gamma_\phi (\theta) = (\cos 2\theta e^{-i\theta} , \sin 2\theta e^{-i\theta} e^{-i\phi}). 	
\end{equation}
It satisfies $\pi(\gamma_\phi)=\alpha_\phi$, with intersections on each fiber forming a 3-design on that fiber. Thus, $\{\gamma_\phi(\theta)\}_{\phi\in[0,\pi]}$ constitutes a family of spherical 3-design curves. Taking half of these  curves,   we get  
\begin{equation}
\widetilde \gamma_\phi (\theta) = (\cos  \theta e^{-i\theta/2} , \sin  \theta e^{-i\theta/2} e^{-i\phi}) ,	
\end{equation}
which corresponds to a family of   unitary 1-design loops $U_\phi(\theta) = R_z(\theta)R_\phi(2\theta) $. We can use Hopf map and stereographic projection to visualize the construction process, as shown in   Fig. \ref{hopf}, more details are given in Appendix \ref{Projection}.

\subsection{Open Unitary 1-design  Path}

So far, we have  considered only closed unitary 1-design paths. We now present   a method for constructing an open unitary 1-design path from a closed   one. Without loss of generality, we take the identity $I$ as  the starting point   and     an arbitrary target operator $U^*$ as the ending point. 
 
Consider, for example, the closed unitary 1-design path $\gamma: s\in [0,1] \mapsto U(\theta=2\pi s) = e^{-i\theta \sigma_z/2}e^{i2\theta \sigma_{\pi/4}/2}$. If $U^*$ does not lie on $\gamma$, there exists an equivalent  unitary 1-design path $WU(\theta)W^\dag$ that passes through $U^*$. To show this, we introduce a  metric $f$ on $ \mathbb{SU}(2)$ defined by   $f(U,V) = |\operatorname{Tr}(UV^\dag)|/2$, which takes values   in $[0,1]$. 
The distance from any point  on $\gamma$   to the   identity is given by
$   f(U(\theta), I) = \left| \cos 2\theta \cos (\theta/4) \right|$, and this expression covers the entire interval $ [0,1]$.   Therefore, for any $U^*$, there exists some  $ \theta^* $  such that $f(U^*, I) = f(U(\theta^*), I)$. Moreover, since   $\operatorname{Tr} U = \operatorname{Tr} V$ and $\operatorname{det} U =  \operatorname{det} V$ imply the existence of   a unitary $W$ such that $U= W VW^\dag$, we may conjugate $\gamma$ by a suitable $W$ to obtain a closed unitary 1-design path containing   $U^*$.

\begin{figure}[t]
\centering
\includegraphics[width=\linewidth]{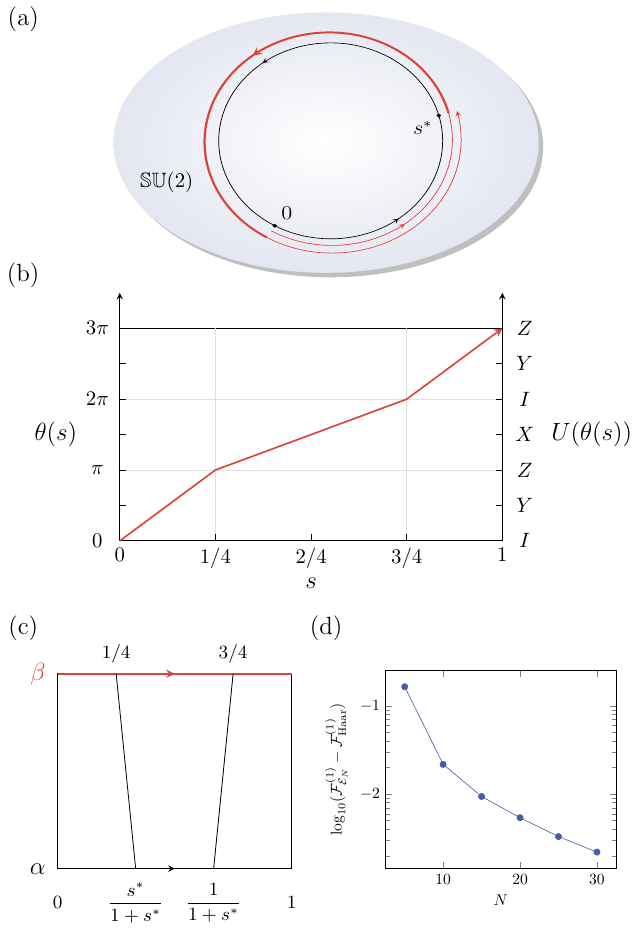}		 
\caption{(a)  Illustration of constructing an open unitary 1-design path from a closed one.  (b) An open unitary 1-design path  connecting $I$ to $Z$ (global phase ignored, and $s^*=1/2$), derived from $U(\theta) = e^{-i\theta \sigma_z/2}e^{i2\theta \sigma_{\pi/4}/2}$ with $\theta$ a piecewise linear function of $s$. (c) From the closed path $\gamma$, we first extract an open path $\alpha$ connecting $I$ to $Z$, and then generate a homotopic, unitary 1-design path $\beta$. (d) Finite  sampling of $\beta$ yields an approximate unitary 1-design, which rapidly converges  to an exact unitary 1-design as the number of sampling points increases.}
\label{open-path}	
\end{figure}

Based on the above analysis, suppose that there exists $s^* \in [0,1]$ such that $U(\theta^* = 2\pi s^*) = U^*$. We then define a path $\alpha(s) = \gamma((1+s^*)s)$ for $s \in [0,1]$, which starts from $I$ and ends at $U^*$. An illustration is provided in  Fig. \ref{open-path}(a). Since $\gamma(s)$ is a closed path, the segment $\gamma([0,s^*])$ is traversed twice by $\alpha$, whereas $\gamma([s^*,1])$ is traversed only once, resulting in nonuniform sampling along $\gamma$.
 To achieve a more uniform coverage, we reparameterize the path to obtain a new one, $\beta(s)$, which can be directly expressed in terms of $\gamma$ as $
\beta(s) = \alpha(\varphi(s))=\gamma\big((1+s^*)\,\varphi(s)\big)$,  
%where $\varphi:[0,1]\to[0,1]$ is a piecewise linear function that maps $[0,s^*/2]$, $[s^*/2,1-s^*/2]$, and $[1-s^*/2,1]$ onto $[0,\, s^*/(1+s^*)]$, $[s^*/(1+s^*),\, 1/(1+s^*)]$, and $[1/(1+s^*),\, 1]$, respectively, thereby reparameterizing the domain of $\alpha$. 
where $\varphi:[0,1]\to[0,1]$ is a piecewise linear reparameterization that adjusts the
relative lengths of the three subintervals $[0,s^*/2]$, $[s^*/2,1-s^*/2]$, and
$[1-s^*/2,1]$ so that they occupy the proportions $s^*/(1+s^*)$, 
$1/(1+s^*) - s^*/(1+s^*)$, and $1 - 1/(1+s^*)$ of the full parameter range for $\alpha$.
This construction redistributes the sampling density along $\gamma$ while preserving the endpoints. The resulting path $\beta$ is homotopic to $\alpha$ and shares the same endpoints:
\begin{numcases}{\beta(s) = }
	\alpha\left(\frac{2s}{1+s^*}\right), &   $s\in[0,s^*/2]$, \nonumber \\
	\alpha\left(\frac{s+s^*/2}{1+s^*} \right), &   $s\in[s^*/2,1-s^*/2]$, \nonumber \\
	\alpha\left(\frac{2s-1+s^*}{1+s^*}\right), &   $s\in[1-s^*/2,1]$. \nonumber
\end{numcases}		
%Given the above analysis, we now suppose that there exists an $s^* \in [0,1]$   such that $U(\theta^*=2\pi s^*) = U^*$. We then consider the path $ \alpha: s\in [0,1] \mapsto  \gamma((1+s^*)s)$, which starts at $I$ and ends at $U^*$. An illustration is provided in  Fig. \ref{open-path}(a). We observe that $\alpha$ traverses  the segment $\gamma([0,s^*])$  twice, but covers $\gamma([s^*,1])$ only once. Intuitively, this imbalance suggests that, to achieve a more uniform distribution, the sampling density along $\gamma([s^*,1])$ should be increased.  To this end, we construct a path that is homotopic to  $\alpha$ and preserves the endpoints, defined by 
%\begin{numcases}{\beta(s) = }
%	\alpha\left(4s \frac{s^*}{1+s^*}\right), &   $s\in[0,1/4]$, \nonumber \\
%	\alpha\left(2s \frac{1-s^*}{1+s^*} + \frac{3s^*-1}{2(1+s^*)} \right), &   $s\in[1/4,3/4]$, \nonumber \\
%	\alpha\left(4s \frac{ s^*}{1+s^*} + \frac{1-3s^*}{2(1+s^*)} \right), &   $s\in[3/4,1]$. \nonumber
%\end{numcases}	
It should be noticed that,  unlike our previous constructions, for this path $\beta$, a   uniform sampling  $\{\beta(k/N)\}_{k=1}^{N }$ actually yields an approximate unitary 1-design for finite $N$, and its distance to an exact unitary 1-design vanishes as $N$ goes large. 

As a concrete example,  Fig. \ref{open-path}(b) shows the explicit form of the   open unitary 1-design path for the target $U^* = Z$, constructed according to the above formula. We sample this path at   equiangular   points to obtain a discrete ensemble of unitaries $\mathcal{E}_N = \{U_k = \beta(k/N)\}_{k=1}^{N}$ and compute the corresponding frame potential $\mathcal{F}_{\mathcal{E}_N}^{(1)} = |\operatorname{Tr}(U_k U_j)|^2/N^2$. It is found that $\mathcal{F}_{\mathcal{E}_N}^{(1)}$ converges exponentially fast in $N$ to the Haar value $\mathcal{F}^{(1)}_\text{Haar}=1$, reflecting  the  unitary 1-design property.

%Under Hopf map, it is mapped to $\alpha(\theta) = (\cos (2\theta), \sin(2\theta), 0)$. For any point $p  $  on the equator, obviously $\alpha(s)$ crosses $p$ twice, i.e.,   there exists some $\theta_p \in [0,\pi]$ such that $p = \alpha(\theta_p) = \alpha(\theta_p + \pi)  $. Since  $ \pi^{-1}(p) =  (\cos(\theta_p) e^{i \phi },  (\sin(\theta_p) e^{i \phi })$, we see that $\gamma(\theta)  \cap \pi^{-1}(p) = \{\gamma(\theta_p) = \beta (\theta_p, \theta_p), \gamma(\theta_p+ \pi) = \beta (\theta_p, \theta_p+\pi)    \}$, corresponding to   antipodal points, and hence a 1-design, of $\pi^{-1}(p) \cong S^1$.
%\[ U(\theta)   =  
%\begin{pmatrix}
%\cos \theta  e^{i  \theta/2} & \sin \theta  e^{i  \theta/2} \\
%-\sin \theta  e^{-i  \theta/2} & \cos \theta  e^{-i  \theta/2}
%\end{pmatrix} = e^{i  \theta \sigma_z/2} e^{i  2\theta \sigma_y/2} \]

\section{Higher-dimensional Systems}
We have presented various constructions of continuous unitary  designs for the qubit system. In this section, we extend these conceptual frameworks and analytical methods to higher-dimensional systems, introducing several approaches for constructing   unitary 1-design paths within the $\mathbb{SU}(d)$ group.

\subsection{Multi-qubit System}
Generalizing the qubit unitary 1-design path to multi-qubit systems is straightforward. 

For two-qubit system, we present a unitary 1-design path as the following 
\begin{equation}
U(\theta) = (e^{-i  \theta \sigma_z/2} e^{i  2\theta \sigma_{\pi/4}/2}) \otimes (e^{-i 4 \theta \sigma_z/2} e^{i  8\theta \sigma_{\pi/4}/2}). 	 \label{2-qubit-1-design}
\end{equation}
To verify this construction, we first establish a general result: Let $  \{U_b \}_{b\in \mathbb{Z}_B}$ be a  $B$-element unitary 1-design in $\mathbb{SU}(2)$, and for each $b$ let there be an associated  $A$-element  unitary 1-design ensemble $  \{U^{(b)}_{a} \}_{a\in \mathbb{Z}_A}$ in $\mathbb{SU}(2)$ (which need not be identical for different $b$).  Then the $AB$-element ensemble $\{U^{(b)}_{a} \otimes U_b\}$ forms a unitary 1-design in $\mathbb{SU}(4)$. This can be demonstrated by examining its action on Pauli   operators:
\begin{align}
	{} & \frac{1}{AB} \sum_{a,b} \left(U^{(b)}_{a} \otimes U_b\right) P_\mu \otimes P_\nu \left(U^{(b)}_{a} \otimes U_b\right)^\dag \nonumber   \\
	 = {} & \frac{1}{B} \sum_{b} \frac{1}{A} \left[\sum_{a} U^{(b)}_{a} P_\mu (U^{(b)}_{a})^\dag \right] \otimes U_b P_\nu U_b^\dag \nonumber \\
	 = {} & \frac{1}{B} \sum_{b}   \frac{  \operatorname{Tr} P_\mu}{2}I   \otimes (U_b P_\nu U_b^\dag) \nonumber \\ 
	 = {} & \frac{\operatorname{Tr} P_\mu  \operatorname{Tr} P_\nu}{4} I  \nonumber \\ 
	 = {} & \frac{\operatorname{Tr} (P_\mu  \otimes P_\nu)}{4} I. \nonumber
\end{align}

Now we use this result  to show that Eq. (\ref{2-qubit-1-design}) defines a 1-design path.
Consider the path at the discrete set   $\{  U(\theta_k) \}$ where $\theta_k =  2\pi k/16  $   for $k \in \mathbb{Z}_{16}$. Decomposing $k= 4a+b$ with $a,b\in \mathbb{Z}_4$, and denoting $R_1 =  e^{-i  \theta_1 \sigma_z/2}$ and $R_2= e^{i  2\theta_1 \sigma_{\pi/4}/2} $ with $\theta_1 = 2\pi/16$, we obtain
\[ U(\theta_k)   = U  ( \frac{2\pi (4a+b)}{16}  ) = \left(R_1^b   P_a R_2^b \right) \otimes P_b,  \nonumber  \]
	in which $P_0=I$, $P_1=iY$, $P_2=iZ$, $P_3=iX$. Clearly, $\{P_b  \}_{b\in \mathbb{Z}_3}$ is a unitary 1-design, and for each $b$, $\{R_1^b   P_a R_2^b  \}_{a\in \mathbb{Z}_3}$ is a 1-design, hence the set $\{U(\theta_k)\}$ constitutes a 1-design.
Next, for any $\varphi\in[0,\pi/16]$, the discrete set $\{  U(\theta_k + \varphi) \}$ is simply the set $\{  U(\theta_k ) \}$ left-multiplied by $e^{-i\varphi \sigma_z/2} \otimes e^{-i4 \varphi \sigma_z/2}$ and right-multiplied by $e^{i2\varphi \sigma_{\pi/4}/2} \otimes e^{i8 \varphi \sigma_{\pi/4}/2}$, and thus   also forms a 1-design. Therefore, we conclude that the   entire path defined in Eq. (\ref{2-qubit-1-design}) constitutes a unitary 1-design path.

This construction generalizes naturally to an  $N$-qubit system as
\begin{equation}
	U(\theta) = \bigotimes_{m=1}^N \left( R_{\bm{n}_{m1}}(4^{N-1}\theta) R_{\bm{n}_{m2}}(2 \times 4^{N-1}\theta)   \right)
\end{equation}
where the rotational axes satisfy $\bm{n}_{m1} \perp \bm{n}_{m2}$ for all $m=1,...,N$.

%\begin{center}
%{\renewcommand{\arraystretch}{1.2}
%\setlength{\tabcolsep}{3pt}
%	\begin{tabular}{|cccc|cccc|}
%	\hline
%	 $U_0$ & $U_4$ & $U_8$ & $U_{12}$ \\ 
%	 $I \otimes I$ & $Y \otimes I$ &  $Z \otimes I$ &  $X \otimes  I$ \\ \hline
%	 $U_1$ & $U_5$ & $U_9$ & $U_{13}$ \\ 
%	  $P I Q \otimes Y$ & $P Y Q \otimes Y$ & $P Z Q \otimes Y$ & $P X Q \otimes Y$ \\ 
%	\hline
%	$U_2$ & $U_6$ & $U_{10}$ & $U_{14}$ \\ 
%	  $P^2 I Q^2 \otimes Z$ & $P^2 Y Q^2 \otimes Z$ & $P^2 Z Q^2 \otimes Z$ & $P^2 X Q^2 \otimes Z$ \\ 
%	\hline
%	$U_3$ & $U_7$ & $U_{11}$ & $U_{15}$ \\ 
%	  $P^3 I Q^3 \otimes X$ & $P^3 Y Q^3 \otimes X$ & $P^3 Z Q^3 \otimes X$ & $P^3 X Q^3 \otimes X$ \\ 
%	\hline
%	\end{tabular}
%}
%\end{center}	

%\begin{widetext}
%\begin{align}
%	\int_0^{2\pi}d\theta U(\theta ) V U^\dag(\theta) & = \sum_{k=0}^{15} \int_{\frac{2\pi k}{16}}^{\frac{2\pi (k+1)}{16}}d\theta U(\theta ) V U^\dag(\theta) \nonumber \\
%	& = \int_{0}^{\frac{2\pi  }{16}}d\theta \sum_{k=0}^{15}  U\left(\theta + \frac{2\pi k}{16} \right) V U^\dag \left(\theta + \frac{2\pi k}{16} \right) \nonumber \\
%	& = \int_{0}^{\frac{2\pi  }{16}}d\theta \sum_{k_2=0}^{3} \sum_{k_1=0}^{3}  U\left(\theta + \frac{2\pi (4k_2+k_1)}{16} \right) V U^\dag \left(\theta + \frac{2\pi (4k_2+k_1)}{16} \right) \nonumber \\
%	& = \int_{0}^{\frac{2\pi  }{16}}d\theta \sum_{k_2=0}^{3} \sum_{k_1=0}^{3}  U\left(\theta + \frac{2\pi (4k_2+k_1)}{16} \right) V U^\dag \left(\theta + \frac{2\pi (4k_2+k_1)}{16} \right)
%\end{align}	
%\end{widetext}

\subsection{ Construction Based on Fiber Bundle Theory}
\label{FiberBundle}
We have seen that for qubit system, unitary 1-designs can be directly  derived   from  spherical designs  due to the isomorphism $\mathbb{SU}(2) \cong \mathbb{S}^3$. For general qudit systems, the situation is more subtle since $\mathbb{SU}(d)$ is not isomorphic to a sphere of dimension $2d-1$. Nevertheless, we can leverage the fiber bundle structure of $\mathbb{SU}(d)$, and spherical designs remain instrumental in constructing unitary designs. In what follows, we first demonstrate this approach for $d=3$ dimensional systems, then proceed to the general case.

Consider the   qutrit system. We analyze    the topological and manifold structure of     $\mathbb{SU}(3)$, basically following   \cite{Khanna97}. $\mathbb{SU}(3)$ is a simply connected, compact,  8-dimensional manifold that is not diffeomorphic to $\mathbb{S }^8$. Indeed, it admits the well-known fibration 
\begin{equation}
	\mathbb{SU}(2) \hookrightarrow \mathbb{SU}(3)  \stackrel{\pi}{\to } \mathbb{S }^5,
\end{equation}
where $\mathbb{SU}(3)$ is the total space, $\mathbb{S }^5$ is the base space and $\mathbb{SU}(2) \cong \mathbb{S}^3$ is the fiber.
 According to the theory of bundles, $\mathbb{SU}(3)$ is the unique (up to isomorphism) nontrivial $\mathbb{SU}(2)$-bundle over $\mathbb{S }^5$. 
 
We now explain the structure in detail. 
Let $\iota$ denote the inclusion of $\mathbb{SU}(2)$ into $\mathbb{SU}(3)$  
\begin{equation}
	\iota : \mathbb{SU}(2) \to \mathbb{SU}(3), \quad 
	\left(\begin{matrix}
	z_1 & z_2 \\
	-\bar z_2 & \bar z_1 
	\end{matrix}\right)   \mapsto 
	\left(\begin{matrix}
	1 &   0 & 0    \\
	0 & z_1 & z_2   \\
	0 & -\bar z_2& \bar z_1
	\end{matrix}\right). \nonumber
\end{equation}
Clearly $\iota ( \mathbb{SU}(2)) \cong \mathbb{SU}(2) $   forms a subgroup of $\mathbb{SU}(3)$. Consider the left action of $\iota (\mathbb{SU}(2))$ on $\mathbb{SU}(3)$, defined by $(W, U) \mapsto WU$ for $ W \in \iota ( \mathbb{SU}(2))$ and $U\in \mathbb{SU}(3)$. This left action preserves the first row of   $U$. Conversely, if two matrices $U_1  $ and $U_2 $ share the same first row, then there   exists   $W  \in \iota ( \mathbb{SU}(2))$ such that $W U_1 = W U_2  $.
Thus, for each $U$ in $\mathbb{SU}(3)$, its right coset    $\{WU: W \in \iota ( \mathbb{SU}(2))\}$  consists of all matrices with the same first row, and each distinct first row corresponds uniquely to a single right coset.  This partitions the entire set $\mathbb{SU}(3)$  into   cosets. We    define the projection $\pi$ mapping each unitary matrix to its first row:
\begin{align}
	\pi: \quad \mathbb{SU}(3) & \to \mathbb{S }^5, \nonumber \\
  U = \left(\begin{matrix}
	c_1 & c_2 & c_3 \\
	* & * & * \\
	* & * & *
	\end{matrix}\right)  & \mapsto c = \left(\begin{matrix}
	  c_1 & c_2 & c_3
	\end{matrix}\right),  \nonumber
\end{align}
where the row vector $c$ satisfies $|c_1|^2 + |c_2|^2 +|c_3|^2  =1$, so $c \in \mathbb{S }^5$. By the preceding analysis, each point in $\mathbb{S }^5$ corresponds to a unique $\mathbb{SU}(2)$ right coset in $\mathbb{SU}(3)$. That is, for each   $c \in \mathbb{S }^5$, the preimage $\pi^{-1}(c)\cong \mathbb{SU}(2)$. Hence, $\mathbb{S}^5$ serves as the base space  and $\mathbb{SU}(2)$ as the fiber.

The next step is   \emph{local trivilization}.
 The fact that $\mathbb{SU}(3)$ is a nontrivial $\mathbb{SU}(2)$-bundle over $\mathbb{S}^5$ means that   $\mathbb{SU}(3)$   looks, locally, like a product $\mathbb{S}^3 \times \mathbb{S}^5$ since $\mathbb{SU}(2) \cong \mathbb{S}^3$, but not globally. The $\mathbb{S}^3$ fiber is  twisted  over the $\mathbb{S}^5$ base in a non-trivial manner. To enable local coordinalization, we consider an open subset of $\mathbb{SU}(3)$ 
\begin{equation}
	\mathcal{M}= \{U \in \mathbb{SU}(3): |c_3| >0 \},    \nonumber
\end{equation}
whose projection is an open subset of $\mathbb{S}^5$
\begin{equation}
	M = \pi(\mathcal{M})=  \left \{c \in \mathbb{S}^5: |c_3| >0  \right\} . \nonumber
\end{equation}
For each point $c \in M \subset \mathbb{S}^5$, we  can select a coset representative  by completing the other   two rows via the Gram-Schmidt process
\[
  \left(\begin{matrix}
	 c_1 & c_2 & c_3
	\end{matrix}\right) \mapsto  \left(\begin{matrix}
		c_1 & c_2 & c_3 \\
		0 & {\displaystyle\frac{ \bar c_3  }{\sqrt{1-|c_1|^2}}} & {\displaystyle \frac{- \bar c_2  }{\sqrt{1-|c_1|^2}}} \\
	\sqrt{1-|c_1|^2}  & {\displaystyle  \frac{- c_2 \bar c_1  }{\sqrt{1-|c_1|^2}}} &  {\displaystyle  \frac{- c_3 \bar c_1  }{\sqrt{1-|c_1|^2}}}  \\ 
	\end{matrix} \right)	.  
\] 
Since $|c_3|>0$,   the denominator  $\sqrt{1 -|c_1|^2}  $   does not vanish, ensuring the matrix is well-defined. This provides a representative element in the coset fiber $\pi^{-1}(c) \subset \mathcal{M}$, and    the full coset can   be obtained by left action of $\iota ( \mathbb{SU}(2) )$ over this element. Thus, any $U \in \mathcal{M}$  is parameterized as:
\[ \left(\begin{matrix}
	1 &   0 & 0    \\
	0 & z_1 & z_2   \\
	0 & -\bar z_2& \bar z_1
	\end{matrix}\right)  \left(\begin{matrix}
		c_1 & c_2 & c_3 \\
	 0 & {\displaystyle\frac{ \bar c_3  }{\sqrt{1-|c_1|^2}}} & {\displaystyle \frac{- \bar c_2  }{\sqrt{1-|c_1|^2}}} \\
	 \sqrt{1-|c_1|^2}  & {\displaystyle  \frac{- c_2 \bar c_1  }{\sqrt{1-|c_1|^2}}} &  {\displaystyle  \frac{- c_3 \bar c_1  }{\sqrt{1-|c_1|^2}}}  \\ 
	\end{matrix} \right). \]
The consequence is that, each unitary   in $ \mathcal{M} \subset \mathbb{SU}(3)$ is  then described by eight local coordinates:     a
complex two-component unit vector $|z_1|^2+|z_2|^2=1$ and a complex three-component unit vector  $|c_1|^2 +|c_2|^2+|c_3|^2 =1$ with $|c_3|>0$.

We   note  that  the complement set  
$ \mathcal{M}^c = \{U \in \mathbb{SU}(3)|  c_3  = 0 \} $ forms a six-dimensional subset of $\mathbb{SU}(3)$ with vanishing measure. Therefore,  $\mathcal{M}$ covers almost full of $\mathbb{SU}(3)$. In this sense, a design set on $\mathcal{M}$ should effectively be  a design set   on $\mathbb{SU}(3)$. 

To construct a 1-design path in $\mathbb{SU}(3)$, we    select a 2-design curve   in the base space $\mathbb{S}^{5}$ and a 2-design point set in $\mathbb{S}^{3}$, then   combine them appropriately. Recall that for   $\mathbb{S}^3$, we have studied the spherical 2-design curve     
$ \xi(\theta)=\frac{1}{\sqrt{2}} (e^{ i \theta}, e^{i2\theta})$. A key property of this curve is that   the discrete set   $\{\xi(2\pi k/5+ \varphi)\}_{k=0}^4$   forms a 2-design point set for any $\varphi \in [0,2\pi/5)$. This motives us to seek  a 2-design curve in $\mathbb{S}^5$ with a periodicity of $5 (2\pi)$. We choose the canonical  2-design curve
\[ \alpha(\theta)=\frac{1}{\sqrt{3}} (1, e^{i  \theta}, e^{i 2\theta}),\]
and lift it  to a curve   $\gamma(\theta)$ in the fiber bundle $\mathcal{M}$
\begin{equation}
\gamma(\theta) = \left\{ \frac{1}{\sqrt{2}} (e^{i \theta}, e^{i 2\theta}),
	\quad \frac{1}{\sqrt{3}} (1, e^{i 5\theta}, e^{i 10\theta})   \right\}, \label{qutrit-1-design} 	
\end{equation}
where $\theta = 2\pi s$ ($s\in [0,1]$). 
As before, this curve satisfies two essential properties:     (i) its projection   runs through $\alpha$ exactly 5 times, satisfying $(\pi \circ \gamma)(s) = \alpha (5s - \left \lfloor 5s \right \rfloor)$, and (ii)  $\gamma$ intersects each fiber exactly five times, with the intersection points forming a 2-design set of the fiber. Specifically, for every    $s$,  we have $\gamma([0,1])\cap \pi^{-1}(\alpha(s)) = \{\gamma(k/5+s):   k=0,...,4\}$.
At last, our constructed unitary 1-design path takes the form
\begin{equation}
	U(\theta) =   \left(\begin{matrix}
	1 & 0 & 0 \\
	0 & z & z^2   \\ 
	0 & - \bar z^2 &   \bar z    \\
	\end{matrix}\right)  \left(\begin{matrix}
		  {\displaystyle \frac{1}{\sqrt{3}}} & {\displaystyle \frac{z^5}{\sqrt{3}}} & {\displaystyle \frac{z^{10}}{\sqrt{3}}} \\
		  0 & {\displaystyle  \frac{ \bar z^{10}}{\sqrt{2}} } & {\displaystyle -\frac{  \bar z^5}{\sqrt{2}} }  \\
		{\displaystyle \frac{\sqrt{2}}{\sqrt{3}}} & {\displaystyle -\frac{  z^5}{\sqrt{6}} } & {\displaystyle -\frac{ z^{10}}{\sqrt{6}} } \\
	\end{matrix} \right),  
\end{equation}
where $z = e^{i\theta}$. The design property can be verified by computing the frame potential. It is found that the path sampled at the $N$ discrete points $\{\theta=2\pi k/N +\varphi\}_{k=0}^{N-1}$ forms a unitary 1-design for any $\varphi\in [0,2\pi/N)$ when $N \ge 25$. 

In general,   the special unitary group $\mathbb{SU}(d)$ possesses a natural principal fiber bundle structure, topologically characterized by an iterative sequence of sphere fibrations. Specifically, $\mathbb{SU}(d)$ forms a locally trivial fiber bundle over the base $\mathbb{S }^{2d -1}$ with fiber   $\mathbb{SU}(d-1)$, a structure arising from the transitive action   of $\mathbb{SU}(d)$ on the unit sphere in $\mathbb{C}^{d}$. This   fibration, 
\[ \mathbb{SU}(d-1)  \hookrightarrow  \mathbb{SU}(d) \to \mathbb{S}^{2d-1} \]
where  $\pi$ maps a unitary matrix to its first row, induces an inductive decomposition. Iterating this construction reveals  $\mathbb{SU}(d)$ as a  twisted product of the odd-dimensional spheres $\mathbb{S}^{2d-1}, \mathbb{S}^{2d-3}, ..., \mathbb{S}^{3}$, fully characterizing its underlying manifold. 

Leveraging this structure,   we construct unitary design paths inductively. Assume we have  $\alpha$ as a unitary design path in $\mathbb{SU}(d-1)$ such that its discrete sampling at $N_{d-1}$   points $\{\theta=2\pi k/N_{d-1} +\varphi\}_{k=0}^{N_{d-1}-1}$ forms a   1-design for any $\varphi\in [0,2\pi/N_{d-1})$ when $N \ge N_{d-1}$. The construction proceeds in two steps: (i) choose the spherical 2-design curve in $\mathbb{S}^{2d-1}$ given by $ (1, e^{i \theta},\ldots, e^{i (d-1)\theta})/\sqrt{d}$; (ii) lift it to the curve $\{\alpha(\theta),  (1, e^{i  N_{d-1} \theta},\ldots, e^{i (d-1)N_{d-1}\theta})/\sqrt{d} \}$. As an example, the unitary  1-design path for $d=3$ in Eq. (\ref{qutrit-1-design}) has $N_3 = 25$, so a
unitary  1-design path for $d=4$ can take the form
\[ \gamma(\theta) = \left\{ \frac{1}{\sqrt{2}} 
\left(\begin{matrix}
	e^{i \theta}\\
	e^{i 2\theta}
\end{matrix}\right),
\frac{1}{\sqrt{3}}
\left(\begin{matrix}
	1\\
	e^{i 5\theta} \\
	e^{i 10\theta}
\end{matrix}\right),
\frac{1}{\sqrt{4}}
\left(\begin{matrix}
	1\\
	e^{i 25\theta} \\
	e^{i 50\theta} \\
	e^{i 75\theta}
\end{matrix}\right) \right\}.
 \]

\subsection{ Construction Based on Heisenberg-Weyl Group}
Our alternative approach is grounded in the well-established framework of the Heisenberg–Weyl (HW) group \cite{Bertlmann08}. For a qudit of dimension $d\ge 2$, we define  the shift operator $X$ and the phase operator $Z$ in the natural basis $\{|j\rangle: j=0,...,d-1\}$  as     
\begin{equation}
X  = \sum_{j=0}^{d-1} |j+1  \;(\bmod\; d)  \rangle \langle j|, \quad Z  = \sum_{j=0}^{d-1} \omega^j |j\rangle \langle j|, \nonumber
\end{equation}
where $\omega = e^{i2\pi/d}$ is the primitive $d$-th root of unity satisfying $\omega^d = 1$. These operators are unitary and obey $X^d = I$ and $Z^d = I$. The HW group is   generated by $X$ and $Z$  as the set of  $d^2$ unitary operators $\{\Lambda_{ab}= X^a Z^b\}$ with $a,b \in \mathbb{Z}_d = \{0,...,d-1\}$. All non-identity elements of this group are traceless.

The HW group generalizes the Pauli group to higher dimensions  and   forms a unitary 1-design.  It has been used, for example, in constructing universal dynamical decoupling sequences \cite{PhysRevLett.134.050601}. Leveraging this group structure, we define a continuous, closed unitary path  as follows:
%\begin{widetext}
%\begin{equation}
%	U(\theta) = W \left(\begin{matrix}
%1 & 0 & \cdots & 0\\
%0 & e^{i   \theta} & \cdots & 0 \\
%\vdots & \vdots & \ddots	 & \vdots \\
%0 & 0 & \cdots & e^{i    (d-1)\theta}
%\end{matrix} \right)  W^\dag \left(\begin{matrix}
%1 & 0 & \cdots & 0\\
%0 & e^{i d \theta} & \cdots & 0 \\
%\vdots & \vdots & \ddots	 & \vdots \\
%0 & 0 & \cdots & e^{i  d (d-1)\theta}
%\end{matrix} \right), \quad \theta\in[0,2\pi],
%\label{qudit-design}
%\end{equation}	
%\end{widetext} 
\begin{equation}
	U(\theta) = W \left[\sum_{k=0}^{d-1} e^{i k\theta}|k\rangle\langle k| \right]  W^\dag \left[\sum_{k=0}^{d-1} e^{i k d\theta}|k\rangle\langle k| \right],
\label{qudit-design}
\end{equation}
where   $W$ is the generalized Walsh-Hadamard transform
%\[ W = \frac{1}{\sqrt{d}} \left(
%\begin{matrix}
%	1 & 1 & 1 & \cdots & 1 \\
%	1 & \omega^{d-1} & \omega^{2(d-1)} &  \cdots &  \omega^{ (d-1)^2} \\
%	1 & \omega^{d-2} & \omega^{2(d-2)} &  \cdots &  \omega^{ (d-1)(d-2)} \\
%	\vdots & \vdots & \vdots & \ddots & \vdots \\
%	1 & \omega & \omega^2 & \cdots & \omega^{d-1} 
%\end{matrix} \right). \]
\[ W = \frac{1}{\sqrt{d}} \sum_{k,j=0}^{d-1} \omega^{-kj}|k\rangle\langle j|, \]
which interchanges the shift and phase operators via the conjugation relation    $W Z^a W^\dag = X^a$ for all $a\in \mathbb{Z}_d$. The qutrit $W$ gate    has been experimentally implemented in, e.g., superconducting system  \cite{PhysRevLett.125.180504}.  One may readily verify that $U(0)=U(2\pi)=0$, confirming that the path is closed.

To demonstrate that the path Eq. (\ref{qudit-design}) yields a unitary 1-design,  we first examine its values at  the discrete set of  equiangular points $\theta_k =  2\pi k/d^2$ for $k \in  \mathbb{Z}_{d^2}$. Decomposing the index $k$   as $k = a \cdot d + b$ with $a,b \in \mathbb{Z}_d$, we write $\theta_k =  2\pi a/d + 2\pi b/d^2$. At these points, the unitary becomes:
\[ U(\theta_k)  =  Q^b  X^a Z^b ,  \]
where
\[ Q =  W\left(\begin{matrix}
1 & 0 & \cdots & 0 \\
0 & e^{i 2\pi  /d^2} & \cdots & 0 \\
\vdots & \vdots & \ddots & \vdots \\
0 & 0 & \cdots & e^{i 2\pi (d-1) /d^2}	
\end{matrix} \right) W^\dag. \]
Thus, the set ${U(\theta_k)}_{k=0}^{d^2 - 1}$ corresponds to HW group elements multiplied on the left by powers of $Q$. We now show that the presence of $Q$ does not affect the 1-design property.

The proof relies on two key properties of   the HW group. First, from the basic commutation relation $\omega XZ =  ZX$, one derives the general relation  $ X^\mu Z^\nu = \omega^{-\mu \nu} Z^\nu X^\mu$, which implies the conjugation rule $\Lambda_{ab} \Lambda_{\mu\nu} \Lambda_{ab}^\dag = \omega^{-a\nu+b\mu} \Lambda_{\mu\nu}$. Second,  the HW group   forms an orthogonal  operator basis. Hence, if the set $ \{U(\theta_k)  \}_{k=0}^{d^2-1}$ averages every  HW operator  except the identity to zero, then it can average  out all traceless Hermitian operators, thus  will be  a unitary 1-design. 

We now verify this condition for an arbitrary non-identity HW operator $\Lambda_{\mu\nu}$ (i.e., $(\mu,\nu) \neq(0,0)$). Consider the  average	 
\begin{align}
  \sum_{k=0}^{d^2-1} U(\theta_k) \Lambda_{\mu\nu} U(\theta_k)^\dag   
& = \sum_{a,b=0}^{d-1} 	Q^b  \Lambda_{ab} \Lambda_{\mu\nu} \Lambda_{ab}^\dag   {Q^b}^\dag \nonumber \\
& = \sum_{a,b=0}^{d-1}  \omega^{-a\nu+b\mu}	Q^b   \Lambda_{\mu\nu}   {Q^b}^\dag. \nonumber
\end{align}
We analyze two possible cases. First, if $\mu \ne 0$ and $\nu=0$, then $\Lambda_{\mu\nu} = X^\mu$, and since $Q$ commutes with $X$, we have 
\begin{align}
\sum_{a,b=0}^{d-1} \omega^{-a\nu+b\mu}	Q^b   X^\mu     {Q^b}^\dag = \sum_{a }^{d-1} \omega^{-a\nu }	   X^\mu  \left[\sum_{ b=0}^{d-1} \omega^{ b\mu}\right]  = 0. \nonumber   
\end{align}
Second, if $\nu \ne 0$,  then  
\begin{align}
\sum_{ b=0}^{d-1}  \omega^{ b\mu}	Q^b   \Lambda_{\mu\nu}   {Q^b}^\dag \left[\sum_{a=0}^{d-1} \omega^{-a\nu}\right] = 0, \nonumber   
\end{align}
From the derivation  we see that, since $Q$ is commutative with $X$ and the summations ensure cancellation, the presence of $Q$ does not alter the result. This confirms that the set ${U(\theta_k)}$ is a unitary 1-design.

We denote $\mathbf{U}_0 = \{U(\theta_k)\}_{k=0}^{d^2-1}$, and define for any $\varphi \in [0, 2\pi/d^2)$   the shifted set $\mathbf{U}_\varphi = {U(\theta_k + \varphi)}{k=0}^{d^2-1}$. 
These sets are related by:
\[ \mathbf{U}_\varphi = \left[ W \sum_{k=0}^{d-1} e^{i k \varphi }|k\rangle \langle k| W^\dag \right] \mathbf{U}_0 \left[\sum_{k=0}^{d-1} e^{i k d \varphi }|k\rangle \langle k| \right]. \]
Since $\mathbf{U}_0$ is a unitary 1-design, and the above relation represents a unitary conjugation and multiplication (which preserve the 1-design property), it follows that $\mathbf{U}_\varphi$ is also a unitary 1-design.  Consequently, the union $\bigcup_{\varphi \in [0, 2\pi/d^2)} \mathbf{U}_\varphi$, which comprises the entire path, forms a unitary 1-design.

\section{Application: Universally Robust Control}

Continuous unitary designs have natural applications in the field of robust quantum control. In the following, we demonstrate that our constructed continuous unitary 1-design paths   provide analytical solutions for achieving universally robust control (URC). 
%We numerically validate their performance by implementing robust single-qubit gates and robust quantum memory, demonstrating their effectiveness under general   noise conditions with unknown amplitudes and arbitrary directions.

\subsection{Theory}
We consider a quantum system with a   Hamiltonian $H_S$ exposed to a   noisy environment. The noise is modeled by a perturbation $V$, characterized by unknown amplitudes along arbitrary   directions in the Liouvillian space. Our objective is to design a time-varying control pulse  $u(t)$ of duration $0\le t \le T$ that generates a control Hamiltonian $H_C(u(t))$ to realize a predefined target gate $\overline U$ while optimally suppressing the noise effects of $V$. Without loss of generality, we omit the identity component   from $H_C$ and $V$    as it only  induces   a global phase, rendering both operators traceless. 

The system's total evolution is   described by  
\[ U_{\text{tot}}(t)=\mathcal{T}\exp\left \{ -i \int_0^t [H_S+H_C(t_1)+V]d t_1\right \},\] 
where $\mathcal{T}$ denotes the time-ordering operator. We assume the system is fully controllable, meaning control solutions exist to realize any target gate in the absence of noise. However, noise causes the actual evolution to deviate from the ideal target, introducing errors. This deviation can be quantified by the gate fidelity
\begin{equation}
F(\overline U,U_{\text{tot}})=
\frac{\left|\operatorname{Tr}[\overline U^\dag U_{\text{tot}}(T)]\right|^2}{d^2},	
\end{equation} 
where $d$ is the system dimension. 
To proceed analytically, we assume that the noise term $V$ is small relative to $H_S$,   allowing us to employ the Dyson perturbative series \cite{Dyson1949}.  
In this framework, the total evolution operator is expanded as  $U_{\text{tot}}(T)=U(T)[I + D^{(1)} + D^{(2)} + \cdots]$, where $U(T)$ is the noise-free evolution operator that implements the target gate,  and 
\[ D^{(k)} = (-i)^k   \!\!\int_{0<t_k<\cdots<t_1<T}\! \prod_{j=1}^k U^\dagger(t_j) V U(t_j) \, dt_j\] 
represents the $k$th-order error term due to presence of $V$.   Truncating the expansion at second order yields an approximate fidelity of   
\begin{equation}
F \approx 1-
\frac{1}{d} \operatorname{Tr} \left[ \left(\int_0^T  U^\dagger(t ) V U(t ) dt  \right)^2  \right].
\label{eq:Fid}
\end{equation}
Hence, the   control  task     amounts to constructing the control evolution $U(t)$ to be of universal  robustness, i.e.,  
\begin{equation}
\int_0^T  U^\dagger(t ) V U(t ) dt    = 0,	\label{robustness-condition}
\end{equation}
for any $V$ in the system's operator  space. This reveals a core requirement for robust control: not only must the control implement the target gate at the final time  $U(T)=\overline U$, but the entire evolution path $U(t)$ must be engineered to dynamically cancel out the noise perturbation throughout the duration of the operation.

Existing robust control strategies are often limited to mitigating noise along specific, pre-characterized directions. Techniques such as dynamical decoupling \cite{dd1,dd2,PhysRevApplied.18.054075}, composite pulses \cite{Levitt86,composite1,composite2,CCP13}, and geometric evolution   optimization \cite{PhysRevLett.111.050404,PhysRevLett.125.250403,PhysRevA.99.052321,PRXQuantum.2.010341,yang2024quantum} can suppress   noise from several directions,  but are generally difficult to generalize to arbitrary system dimensions and target gates. Conversely, while quantum error correction \cite{PhysRevLett.84.2525,devitt2013quantum} and feedback control \cite{brif2010control,zhang2017quantum} can manage general noise, they incur substantial resource overhead, requiring either auxiliary qubits or adaptive measurements. An alternative approach employs multi-objective optimization  to  suppress multiple, coexisting noise channels concurrently \cite{Cory19,PhysRevApplied.21.034042}. However, these numerical methods are often hampered by high computational complexity and a strong dependence on problem-specific parameters.  

The recent work in Ref.~\cite{Poggi2024} addresses universally robust control (URC) directly. The authors recognize that the universally robust condition in Eq. (\ref{robustness-condition}) is equivalent to requiring the noise-free evolution   $\{U(t)\}_{t=0}^T$ to form a unitary 1-design. To obtain such a unitary 1-design control, they employ numerical means. The total evolution is discretized into $N$ segments of equal duration, and a control pulse is numerically optimized to render the resulting set of operators $\{U(t_k)\}_{k=0}^{N-1}$   a unitary 1-design. While effective for small systems, this approach faces   computational bottlenecks for multi-qubit systems due to the exponential growth of the Hilbert space dimension. Furthermore, such numerical solutions often lack physical interpretability. These limitations highlight the need for analytical solutions to the URC problem.

We now employ the continuous unitary 1-designs constructed in the previous section to solve the URC problem. By definition, a continuous unitary 1-design $U(s)$ is a smooth path in   $\mathbb{SU}(d)$ that satisfies 
\begin{equation*}
	\int_{[0,1]} U^\dag(s) V U(s)\, ds 
	= \int_{\mathbb{SU}(d)} U^\dag V U \, dU 
	= \frac{I}{d}\operatorname{Tr}(V) =  0,
\end{equation*}
for any traceless operator $V$, fulfilling the universally robust condition. Consequently,  by inversely engineering the control fields from a predefined continuous unitary 1-design path, we obtain a direct and constructive method for synthesizing smooth URC fields. The following numerical simulations demonstrate the implementation and performance of this approach.

\begin{figure}[b]
    \centering
    \includegraphics[width=\linewidth]{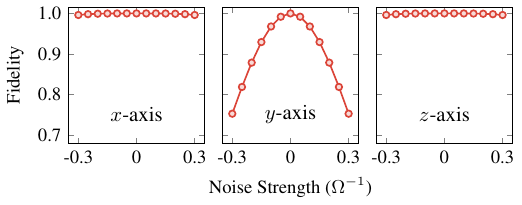}
    \caption{A rectangular $2\pi$ rotation  pulse about the $y$-axis   generates  a  symmetric continuous unitary 1-design path capable of averaging out $X$ and $Z$ noise operators. This makes the pulse robust against noise with $x$- and $z$-components, but it remains highly susceptible to noise along the $y$-axis.}
    \label{figure3}
\end{figure}

\subsection{Analytic Control Solution}
For our numerical analysis, we focus on the single-qubit system. We assume resonant controls  that act along all three axes, described by the vector  $\bm {u}(t) = \big(u_x(t), u_y(t), u_z(t)\big)$ for $t \in [0,T]$, with a maximum Rabi frequency $\Omega$ such that $|\bm {u}(t)| \leq \Omega$. This control model is experimentally implementable. For instance, superconducting qubits utilize XY microwave drives and Z flux biases \cite{Oliver19}. In   the resonant frame,  the  control Hamiltonian is 
\begin{equation}
	H(t)=u_x(t) \frac{\sigma_x}{2} +u_y(t) \frac{\sigma_y}{2}  +u_z(t) \frac{\sigma_z}{2}. \label{H(t)}
\end{equation}
We employ the  unitary 1-design path $U(\theta) = R_{\bm{n}_1}(\theta) R_{\bm{n}_2}(2\theta)$  from Eq.~(\ref{1-design}). It is important to note that this path is not realized by sequentially applying two   gates. Rather, it represents a     path parameterized by a path parameter $s\in[0,1]$, which must be generated through continuously controlled evolution with respect to the physical time parameter $t$.   We   derive the corresponding URC field  by solving  the Schr{\"o}dinger equation $\dot U(\theta(t)) = -i H(t) U(\theta(t))$, yielding
\begin{align}\label{h2}
H(t) & = i \dot U(\theta) U^\dag(\theta) \nonumber \\
	&=\frac{d\theta}{d t}\left[ \frac{1}{2} (\bm{n}_1 \cdot \bm{\sigma}) + R_{\bm{n}_1} (\theta)(\bm{n}_2 \cdot \bm{\sigma}) R^\dag_{\bm{n}_1} (\theta)\right] \nonumber \\
	&=\frac{d\theta}{d t}\left[ \frac{1}{2} (\bm{n}_1 \cdot \bm{\sigma}) + [\cos\theta \bm{n}_2 +\sin\theta (\bm{n}_1 \times \bm{n}_2)] \cdot \bm{\sigma}\right], \nonumber
\end{align}
where we have used Pauli matrix identity   $(\bm{n}_1 \cdot \bm{\sigma})(\bm{n}_2 \cdot \bm{\sigma})=(\bm{n}_1 \cdot \bm{n}_2) I+i (\bm{n}_1 \times \bm{n}_2) \cdot \bm{\sigma}$.
Comparing this with the Hamiltonian in Eq. (\ref{H(t)}), we identify the control field
\begin{equation}
	\bm {u}=\frac{d\theta}{ d t} [\bm{n}_1+ 2 \cos\theta \bm{n}_2+ 2 \sin\theta (\bm{n}_1 \times \bm{n}_2)].
	\label{u}
\end{equation}
This control field must satisfy the power constraint  $|\bm{u}(t)|\le  \Omega$, which amounts to the requirement $|d\theta/dt| \le \Omega/\sqrt{5}$.  
In the case where $\theta$ is linearly parameterized as $\theta = 2\pi s$,    the physical time $t$ can be related to the path parameter $s$ via $s = t\Omega / (2\pi \sqrt{5})$.

\begin{figure*} 
    \centering
    \includegraphics[width=0.96\linewidth]{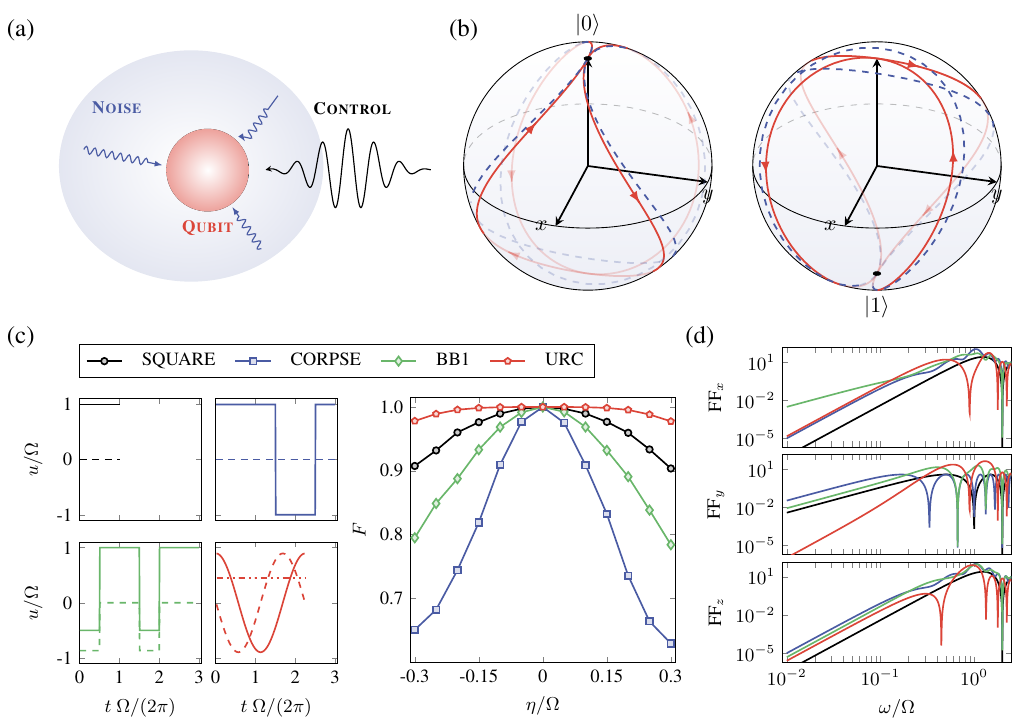} 
    \caption{Simulation results for implementing a single-qubit identity gate against an unknown noise environment using different control pulses. (a) Problem setup. A single qubit   is subject  to a generic  noise perturbation with an unknown amplitude  and orientation.  The goal is to use active control to protect the qubit from this noise,   thereby implementing an identity gate.
    (b) State evolution trajectories  under   URC pulse control, starting from the $|0\rangle$ state (left) or the $|1 \rangle$ state (right). The solid curve represents the ideal evolution path (a state 1-design curve). The dashed curve shows the evolution when noise is present. Although the noisy evolution   deviates from the ideal one, it automatically corrects the errors by the end of the evolution, forming a dynamically corrected gate.
    (c) Control waveforms and static noise robustness. The plots show the control waveforms for the different pulses. Dashed and solid lines represent the $x$- and $y$-components,  respectively. For the URC pulse, a constant $z$-component is also present (dash-dotted line). All pulses are bounded by a maximum Rabi frequency $\Omega$. The graph on the right compares the robustness of each pulse to static noise, showing the noise averaged gate fidelity $F_N$ as a function of   noise strength $\eta $. For each   $\eta $, the noise-averaged fidelity $F_N$ is computed from an average over 1000 random spatial orientations of the noise vector. 
    (d) Robustness to time-varying noise. The filter functions $\mathrm{FF}_\alpha$ ($\alpha = x,y,z$) for each pulse, quantifying their susceptibility to time-dependent noise along different spatial axes. } 
    \label{figure7}
\end{figure*}

\subsection{Numerical Simulations}
\subsubsection{Robust quantum gate}

We first   consider the implementation of a robust single-qubit identity gate,  $\overline U = I$. In the presence of noise, a qubit's state becomes unstable. Applying active control to counteract these noise effects ensures the net evolution is the identity, thereby preserving quantum information. A robust identity gate is also useful in multi-qubit circuits for selectively decoupling and preserving the state of local qubits.

For noise   restricted to one or two specific directions, we note that a simple square pulse is sufficient to achieve first-order decoupling. This is interpretable within the framework of symmetric unitary  designs \cite{PRXQuantum.4.040331}. Consider, for instance, a noise operator $V = V_x \sigma_x/2 + V_z \sigma_z/2$ with nonzero $x$- and $z$- components.  The discrete   set $\{I,Y\}$   averages this noise to zero, since $IVI + YVY=0$. This occurs because  $\{I,Y\}$ constitutes a  symmetric unitary 1-design for the $Y$-symmetric subgroup $\{U\in \mathbb{SU}(2)| [U,Y]=0 \}$, which is isomorphic to $\{R_y(\theta)| \theta \in \mathbb{R}\} $. Obviously, a continuous unitary 1-design path that connects the elements $\{I,Y\}$ is the subgroup $\{R_y(\theta)\} $ itself. The corresponding control field for this path is simply a constant square pulse effecting a $2\pi$ rotation about the $y$-axis.

However, if the noise has components along all three axes, a primitive square pulse about the $y$-axis becomes ineffective at suppressing its $y$-component, as illustrated in Fig. \ref{figure7}(a). In this more general case, finding a robust control is not as straightforward.  Our continuous unitary design theory provides a solution, given by Eq. (\ref{u}). Specifically, by choosing $\bm{n}_1=(0,0,1)$, $\bm{n}_2=(0,1,0)$, and $\theta=2\pi s= \Omega t$, the URC pulse takes the form 
%\begin{equation}
%	u_x(t)  =-\frac{2\Omega}{\sqrt{5}}\sin(\Omega t), u_y(t)  =\frac{2\Omega}{\sqrt{5}}\cos(\Omega t), u_z(t)=\frac{\Omega}{\sqrt{5}}
%\end{equation}
\begin{equation}
	 \left\{
      \begin{aligned}
        u_x(t) &= -\frac{2\Omega}{\sqrt{5}}\sin\left(\frac{ \Omega}{\sqrt{5}} t\right), \\
        u_y(t) &=  \frac{2\Omega}{\sqrt{5}}\cos\left(\frac{ \Omega}{\sqrt{5}} t\right), \\
        u_z(t) &= \frac{\Omega}{\sqrt{5}}, \\
      \end{aligned}
    \right.
\end{equation}
 with $t\in[0,2\pi \sqrt{5} /\Omega]$. This pulse naturally implements a universally robust identity gate.
For comparison, we also evaluate   the widely used composite pulses CORPSE \cite{PhysRevA.67.042308} and BB1 \cite{PhysRevA.70.052318}, which are   designed to mitigate static detuning   and amplitude noise, respectively. CORPSE consists of three elementary gates $R_{\phi_3}(\theta_3)R_{\phi_2}(\theta_2)R_{\phi_1}(\theta_1)$, where $\theta_1=3\pi,\theta_2=2\pi,\theta_3=\pi$ and $\phi_1=\phi_3=\pi/2,\phi_2=3\pi/2$. BB1 consists of four elementary gates $R_{\phi'_4}(\theta'_4)R_{\phi'_3}(\theta'_3)R_{\phi'_2}(\theta'_2)R_{\phi'_1}(\theta'_1)$, where $\theta'_1=\theta'_3=\pi,\theta'_2=\theta'_4=2\pi$ and $\phi'_1=\phi'_3=7\pi/6, \phi'_2=5\pi/2,\phi'_4=\pi/2$.
The pulse waveforms for these sequences, along with the primitive square pulse, are shown in Fig.~\ref{figure7}(c). We compare their performance by simulating the gate fidelity $F$ under static, randomly oriented three-dimensional noise with strength  $\eta$ ranging over $[0, 30\% \Omega]$, as also shown in Fig.~\ref{figure7}(c). The results clearly show that both CORPSE and BB1 perform worse than the square pulse in this scenario.   This is because they are designed to counteract a specific type of noise. For example, while  CORPSE provides better robustness than SUQARE against  detuning noise  along $z$-axis, its average performance is lower when considering arbitrary noise orientations.  In contrast, our URC pulse demonstrates superior, universal robustness, maintaining high gate fidelity across the entire range of tested noise. For visualization, in Fig.~\ref{figure7}(b), we plotted the state evolution under URC pulse control. Applying a unitary 1-design to a quantum state  generates a state 1-design. We observe that  the state evolution trajectories are, as expected, evenly distributed on the Bloch sphere. When noise is present, the evolution deviates from its original path but eventually arrives at the target state, with errors being automatically corrected.

Furthermore, we investigate the robustness of the URC pulse against time-varying noise, where the noise amplitude fluctuates during the pulse execution. Previous research has shown that composite pulses can maintain robustness against time-dependent, non-Markovian noise at frequencies up to 10\% of the Rabi frequency \cite{PhysRevA.90.012316}. We therefore anticipate that our URC pulse possesses a similar capacity to mitigate low-frequency noise to a certain extent. To verify this, we employ the filter function  (FF) formalism \cite{Biercuk13}, which quantifies a control pulse's sensitivity to noise across different frequencies and thereby characterizes its impact on gate fidelity. The FF is typically defined as 
\begin{equation}
	\text{FF}_\alpha(\omega)=\sum_{k} \left| -i\omega \int_0^T R_{\alpha k}(t) e^{i\omega t} dt \right|^2, 
\end{equation} 
where $R_{\alpha k}(t)=\text{Tr}[U^\dag(t)\sigma_\alpha U(t)\sigma_k]/2$ and $\alpha, k=x,y,z$. Figure \ref{figure7}(d) compares the FFs along different directions for the primitive $y$-rotation, CORPSE,  BB1, and URC pulses as shown in Fig. \ref{figure7}(c). 
We find that for low-frequency time-dependent noise along $y$-direction,   the  $\operatorname{FF}_y$ of the URC pulse is significantly smaller than that of all other pulses, confirming its superior robustness.  Furthermore, we observe that the URC pulse exhibits filter functions of comparable magnitude along all three   directions. In contrast, the other pulses only show markedly reduced FFs along   one or two specific axes. This result underscores the URC pulse's unique ability to suppress time-dependent noise of arbitrary orientation.

\begin{figure}
    \centering
    \includegraphics[width=\linewidth]{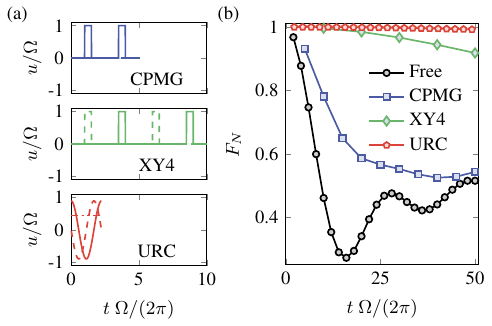}
    \caption{Simulation results for quantum memory based on different DD sequences. (A) Control waveforms for each sequence,    all subject to the power constraint $|\bm{u}| \le \Omega$.  (B)     Decay of noise-averaged fidelity versus the number of sequence repetitions.  
    The total memory time is identical for all sequences, requiring a different number of repetitions for each.}
    \label{figure4}
\end{figure} 

\subsubsection{Quantum memory}
Next, we   explore the application of continuous unitary 1-designs to qubit quantum memory. Conventionally, quantum memory is protected by dynamical decoupling, which preserves quantum coherence from environmental noise by repeatedly   flipping the  qubit so that  the qubit-environmental interaction is averaged out \cite{RevModPhys.88.041001}.
 As demonstrated above, the URC pulse implements a robust identity gate, making it a promising candidate for quantum memory through repeated application. For comparison, we also  evaluate standard dynamical decoupling sequences, including  CPMG and XY4. The CPMG  sequence  \cite{meiboom1958modified},  denoted as $f_\tau R_y(\pi) f_{2\tau} R_y(\pi) f_\tau$, where $f_\tau$ represents free evolution for a duration $\tau$, is designed to  suppress noise  perpendicular to its   pulse rotation axis. The XY4 sequence \cite{gullion1990new}, taking the form $f_\tau R_y(\pi) f_{2\tau} R_x(\pi) f_{2\tau} R_y(\pi) f_{2\tau} R_x(\pi) f_\tau$, is engineered to mitigate noise along all three spatial directions.
The explicit pulse waveforms for these sequences are shown in Fig.~\ref{figure4}(a).  

During each sequence, a noise perturbation with a strength up to $0.05\Omega$ is applied along a random three-dimensional orientation. Each sequence (including free evolution periods) is repeated multiple times to generate a fidelity decay curve.  Averaging over 100 noise realizations of different random directions and random strengths then gives the noise-averaged fidelity $F_N$,  as shown in Fig.~\ref{figure4}(b). As expected, we find that XY4 outperforms CPMG, since XY4 is designed to suppress noise along all three spatial directions, while CPMG only addresses two. The XY4 sequence is derived from the Pauli group $\{I,X,Y,Z\}$, which forms a unitary 1-design.  Ideally, this design should enable the sequence to have universal decoupling ability.  In practice, however, the finite duration of its
$\pi$ pulses, causes deviations from ideal instantaneous rotations, compromising its robustness against fully generic, three-dimensional noise. Remarkably, the URC sequence preserves the characteristics of a unitary 1-design throughout its continuous evolution. As a result, it maintains near-perfect coherence over the investigated time range, highlighting its significant potential for realizing reliable quantum memory in universal noise environments.

\section{Discussions and Outlook}
Our work   extends  the scope of unitary design research from the discrete to the continuous domain. As an initial work, we  have developed theoretical methods for constructing continuous unitary designs,    provided  explicit construction examples, and demonstrated their practical applications in  quantum control.

We first dedicated   considerable effort to constructing single-qubit unitary 1-design paths and elucidating their geometric and topological principles. Indeed,   even for   the   single-qubit case, this task is highly non-trivial. Unitary designs basically aim to uniformly sample the unitary group  $\mathbb{SU}(2)$. A naive approach is to impose a coordinate system that treats the group manifold $\mathbb{SU}(2)$ as a patch of $\mathbb{R}^3$. However, many  of such coordinate systems introduce metric distortions   relative to the intrinsic geometry of the $\mathbb{SU}(2)$. For example, random sampling of Euler angles $(\alpha, \beta, \gamma)$ in the $ZYZ$ decomposition $U = R_z(\alpha)R_y(\beta)R_z(\gamma)$ (up to a global phase) leads to   significant redundant sampling.   This method is particularly ill-suited for   continuous paths, as it is prone to singularities \cite{stuelpnagel1964parametrization} and gimbal lock \cite{shoemake1985animating}.
%, a pathological degeneracy where two rotational axes align and a degree of freedom is lost.
The inefficiency comes from topological mismatch: the Euler angle representation has the topology of $\mathbb{S}^1\times \mathbb{S}^1 \times \mathbb{S}^1$, which fails to capture the true topology of $\mathbb{SU}(2)$. In contrast, the Hopf fibration   correctly captures this topology.
    Consequently, Hopf coordinates naturally describe the intrinsic structure of $\mathbb{SU}(2)$ and provide a   tool for generating uniform distributions \cite{Mitchell10},  explaining their essential role in constructing unitary design paths \cite{Lindblad1}.    

We also constructed open unitary 1-design paths by strategically modifying closed ones. The presented examples are approximate unitary designs when discretely sampled, but converge to exact designs with increasing sampling density. This limitation can be mitigated by a more sophisticated parameterization, for instance, by assigning a local weight to each segment and reparameterizing the path to make the sampling density inversely proportional to this weight, thereby ensuring adaptive and efficient coverage. These open unitary 1-design paths will find applications in dynamically corrected gates \cite{PhysRevLett.102.080501} and dynamical-decoupling–protected gates \cite{PhysRevLett.105.230503,PhysRevLett.112.050502}, as they simultaneously implement a target quantum gate while averaging out noise-induced errors.

For higher-dimensional systems, we have presented two distinct   methods for constructing unitary 1-design paths: one based on fiber bundle theory and the other on the Heisenberg-Weyl group. These methods yield paths of different lengths.  The Heisenberg-Weyl approach  draws upon a well-established framework in quantum information \cite{vourdas2004quantum,wang2020qudits}, bearing a group-theoretic flavour. In contrast, the fiber bundle approach offers a complementary geometric perspective. This mathematical framework describes spaces that are locally simple yet may exhibit complex global topological twists. By revealing the underlying geometric structure of the unitary design paths, it brings powerful topological tools into play, thereby broadening the methodological landscape beyond traditional group-theoretic constructions.

Finally, we demonstrated the practical utility of our unitary 1-design paths by employing them to generate universally robust control pulses. These pulses achieve high-fidelity identity operations under arbitrary unknown static noise, as confirmed by numerical comparisons with conventional techniques. Furthermore, filter function analysis verifies their robustness against slowly varying noise. While the current sequences do not yet account for common experimental errors like Rabi inaccuracies \cite{PhysRevA.82.042306,PhysRevLett.118.133202}, this limitation can be directly addressed by incorporating time-symmetry between consecutive sequences \cite{RevModPhys.76.1037,brinkmann2016introduction}. Moreover, our framework for constructing qudit unitary design paths can be directly applied to implement qudit dynamical decoupling \cite{PhysRevLett.134.050601,PhysRevResearch.3.013235}. Given the persistent challenge of designing robust pulses in noisy environments, our work provides a valuable addition to the standard toolbox for robust quantum engineering.

Our study of continuous unitary designs paves the way for numerous promising research directions. Below, we outline several open questions that merit further investigation.
\begin{enumerate}[itemsep=0pt]
	\item[(1)] How to find  optimal, or asymptotically optimal, continuous unitary 1-design paths? This problem, which mirrors the classical pursuit of optimal spherical designs \cite{BRV13,Ehler2023,Lindblad1}, is of significant theoretical interest. On practical side,    the shortest  such  path may lead to time-optimal control solutions in quantum control. While our work has presented several explicit unitary 1-design paths of varying lengths, we find that especially those constructed via the inductive procedure based on topological bundle theory (as described in Sec. \ref{FiberBundle}) exhibit a length that grows rapidly with the system dimension.  We anticipate that shorter paths exist in abundance; identifying them warrants further exploration and may involve techniques from other mathematical theories, such as additive combinatorics \cite{Bannai09,Bajnok18,Bajnok24}.	
	\item[(2)] This work focuses on continuous unitary 1-designs ($t=1$) and their applications in quantum control. A natural next step is to construct continuous unitary designs for $t \ge 2$ and explore their potential applications. For $t=2$, the most straightforward approach is to find a continuous path that passes through a discrete unitary 2-design set, such as the Clifford group, such that the path itself forms a continuous 2-design. In the single-qubit case, based on the correspondence between $\mathbb{SU}(2)$ and   $\mathbb{S}^3$, a unitary 2-design can be derived from a spherical 4-design. The latter can be constructed using the Hopf map technique described in Ref. \cite{Lindblad2}, which involves placing a spherical $4$-design on each fiber of the Hopf map associated with a $2$-design curve on $\mathbb{S}^2$. For higher-dimensional systems, the construction might become    more challenging. Given a successful construction, it would be highly interesting to identify applications. Unitary $2$-designs are crucial for many quantum information protocols, such as randomized benchmarking \cite{Laflamme07,PhysRevA.80.012304}, suggesting immediate practical utility.
    \item[(3)] Continuous unitary designs can be integrated into  Hamiltonian engineering as they enable   controls over average Hamiltonians. There exist connections between unitary designs and decoupling sequences. A unitary 1-design, for example, serves as a universal decoupling group for its ability to average any traceless Hamiltonian to zero \cite{PhysRevLett.134.050601}. A more refined Hamiltonian decoupling can possibly be enabled by symmetric unitary designs \cite{PRXQuantum.4.040331}. Usually,   decoupling sequences employ discrete, instantaneous $\pi$ and $\pi/2$ pulses. We are thus to ask: can continuous pulse policies, derived from symmetric unitary designs, achieve the same decoupling goal? More broadly, analogous to the role of discrete pulse sequences in average Hamiltonian theory for engineering many-body spin dynamics \cite{PhysRevX.10.031002},   continuous unitary designs   may provide a continuous framework for averaging out undesired interactions, thereby advancing the toolkit for Hamiltonian engineering.
\end{enumerate}
In summary, this work establishes a foundation for several future research avenues. On the theoretical front, the further exploration of continuous unitary design construction promises to uncover new examples and stimulate the development of associated mathematical tools.   For applications, these designs provide a practical tool for Hamiltonian engineering, directly advancing capabilities in quantum decoupling, sensing, and simulation.

\section{Acknowledgments}
This work is supported by   the Innovation Program for Quantum Science and Technology (2024ZD0300400), the National Natural Science Foundation of China (grant no. 12441502, 12204230), and the Guangdong Provincial Quantum Science Strategic Initiative (GDZX2305006, GDZX2405002).

\appendix

\section{Continuous Spherical 2-design and Unitary 1-design for Qubit}
\label{appendix_A}
Let $V$ be a  traceless Hermitian operator. In the Pauli basis, $V$ admits the   decomposition    $V = \bm{v}\cdot \bm{\sigma}$, where $\bm{v}\in \mathbb{R}^3$. Let $U $ be a unitary matrix parameterized as
\[ U = \left(\begin{matrix}
	x_1 + i x_2 & x_3 + i x_4 \\
	-x_3 + i x_4 & x_1 - i x_2
\end{matrix}\right),   \]
with $(x_1,x_2,x_3,x_4) \in \mathbb{S}^3 \subset \mathbb{R}^4$. This establishes the known identification of $\mathbb{SU}(2) \cong \mathbb{S}^3$.
The conjugation $W = UVU'$ preserves both the Hermitian and traceless properties of $V$, so it can be expressed as   $W  = \bm{w}\cdot \bm{\sigma}$ for some $\bm{w}\in \mathbb{R}^3$. A direct computation shows   that each   component   $w_k$  is a homogeneous  quadratic form in $x$, specifically, $ w_{k} =  x A_{k} x^T $,  
where the symmetric matrices $A_k$ are given by
\begin{align}
A_{1} & = \left(\begin{matrix}
	 v_1 & v_2 & -v_3 &  0 \\
	  v_2 & -v_1 & 0 & v_3\\
	 -v_3  & 0 & -v_1 & v_2 \\
	 0  & v_3 & v_2 &   v_1 	\end{matrix}\right), \nonumber \\
A_{2} & = \left(\begin{matrix}
	 v_2 & -v_1 & 0 &  v_3 \\
	  -v_1 & -v_2 & v_3 & 0\\
	 0  & v_3 & -v_2 & v_1 \\
	  v_3  & 0 & v_1 &  -v_2 	\end{matrix}\right), \nonumber \\
A_{3} & = \left(\begin{matrix}
	 v_3 & 0 & v_1 &  -v_2 \\
	  0 & v_3 & v_2 & v_1\\
	 v_1  & v_2 & -v_3 & 0 \\
	 -v_2  & v_1 & 0 &  -v_3 	\end{matrix}\right). \nonumber 	 
\end{align}  
On the 3-sphere $\mathbb{S}^3$,  the following integration identity holds for the coordinate functions: $\int_{\mathbb{S}^3} x_k x_j d\mu = \delta_{kj}/4$. Suppose $ \gamma  $ is a spherical 2-design curve in $\mathbb{S}^3$, meaning that for any polynomial $f$ of degree at most 2, there is $\frac{1}{\ell(\gamma)}\int_\gamma f  ds = \int_{\mathbb{S}^3} f d\mu$, so
\[ \frac{1}{\ell(\gamma)} \int_{\gamma} w_k ds = \int_{\mathbb{S}^3} w_k d\mu = \frac{1}{4}\operatorname{Tr}A_k = 0.\]
Now, let $U_\gamma$ be the unitary path in $\mathbb{SU}(2)$ corresponding to $\gamma$,  we thence have 
\[
	\frac{1}{\ell(\gamma)}  \int_{U_\gamma} UVU'dU   = \frac{1}{\ell(\gamma)} \int_{\gamma} \bm{w}\cdot \bm{\sigma} ds = 0, \] 
 which holds for all traceless Hermitian  $V$. This implies that $U_\gamma$ 
  is a unitary 1-design path.

\section{Stereographic Projection} 
\label{Projection}
Stereographic projection provides a powerful method for visualizing the geometry of the 3-sphere $\mathbb{S}^3$ by mapping it onto $\mathbb{R}^3$. We employ this technique to visualize the   design curves on $\mathbb{S}^3$.

Consider the Hopf fibration $\mathbb{S}^1  \hookrightarrow  \mathbb{S}^3 \stackrel{\pi}{\to } \mathbb{S}^2$, where $\pi$ is Hopf map. Let $C$ denote the preimage of the   great circle $(0, \cos \theta, \sin \theta)$ in the base   $\mathbb{S}^2$. This preimage is a circle of circles (fibers), namely a torus, named   the Clifford torus, having the explicit  formula
\[ C(\theta,\phi) =  \frac{1}{\sqrt{2}} (\cos \theta, \sin \theta, \cos \phi, \sin \phi),   \] 
where   $\theta,\phi \in [0,2\pi]$.    To visualize  this geometric object  embedded in $\mathbb{S}^3$, we employ the stereographic projection $f: \mathbb{S}^3 \setminus {(0,0,0,-1)} \to \mathbb{R}^3$ from the south pole $(0,0,0,-1)$ to the equatorial plane $x_4 = 0$, defined by 
\[ f(x_1,x_2,x_3,x_4) = \frac{1}{1+x_4} (x_1,x_2,x_3). \]
This conformal mapping preserves angles and local shapes, establishing a homeomorphism between the punctured sphere and the plane. Under this projection,  the Clifford torus transforms to
\[ f(C) = \frac{1}{\sqrt{2}+\sin \phi}(\cos \theta, \sin \theta, \cos \phi), \]
revealing its structure in three dimensional space.
The  spherical design curves $\xi$ in Eq. (\ref{xi}) and $\gamma$ in Eq. (\ref{gamma}) both reside on this   torus, projecting to
\begin{align}
	f(\xi) & = \frac{1}{\sqrt{2}+\sin 2\theta}(\cos \theta, \sin \theta, \cos 2\theta) \nonumber \\
	f(\gamma) & = \frac{1}{\sqrt{2}-\sin 3\theta}(\cos \theta, \sin \theta, \cos 3\theta) \nonumber
\end{align}
The design curves $\xi'$ in Eq. (\ref{xip}) and $\gamma'$ in Eq. (\ref{gammap}) are obtained from $\xi$ and $\gamma$, respectively, through application of the same four-dimensional rotation $R$.  These rotated curves reside on a different torus in $\mathbb{S}^3$, specifically the Clifford torus transformed by the rotation 
\begin{equation}
	R(C) = (\cos \theta \cos \phi, \cos \theta \sin \phi, \sin \theta \cos \phi, \sin \theta \sin \phi),  \nonumber
\end{equation}
where   $\theta,\phi \in [0,2\pi]$.
By applying a corresponding rotation to both the center of projection and the projection plane, we maintain the same visualization characteristics as in the unrotated case.
%We choose the  projection plane to be the equatorial plane of the torus namely $C(\theta, \phi=0)$, and the center of projection to be a point perpendicular to this plane, say the north pole  $q = (0,1,1,0)/\sqrt{2} \in \mathbb{S}^3$. 

%\bibliography{manuscript}

\end{document}